\documentclass[onecolumn,11pt]{article}
\usepackage{mathrsfs}
\usepackage{amsmath, alltt}
\usepackage{rotating}
\usepackage{graphicx}
\usepackage{subfigure}
\usepackage[onehalfspacing]{setspace}
\usepackage[hmargin={0.5in, 0.5in}, vmargin={0.8in, 0.8in}]{geometry}
\usepackage{amsfonts}
\usepackage{amssymb}
\usepackage{natbib}
\usepackage{amsmath}
\usepackage{bm}
\usepackage{latexsym}
\usepackage{amssymb,bm}
\usepackage[section]{placeins}
\usepackage{longtable}
\usepackage{lscape}
\usepackage{scalefnt}
\usepackage{setspace}
\usepackage{color}
\usepackage{hyperref}

\DeclareMathOperator*{\argmax}{arg\,max}

\begin{document}


\title{A new perspective from a Dirichlet model for  forecasting outstanding liabilities of nonlife insurers}

\author{
    Karthik Sriram \\
    Indian Institute of Management Ahmedabad \\
    Email: karthiks@iima.ac.in \\
    \and
    Peng Shi \\
    Wisconsin School of Business\\
    University of Wisconsin-Madison\\
    Email: pshi@bus.wisc.edu\\
}

\maketitle
\abstract{
Forecasting the outstanding claim liabilities to set adequate reserves is critical for a nonlife insurer's solvency. Chain-Ladder and Bornhuetter-Ferguson are two prominent actuarial approaches used for this task.  The selection between the two approaches is often ad hoc due to different underlying assumptions. We introduce a Dirichlet model that provides a common statistical framework for the two approaches, with some appealing properties. Depending on the type of information available, the model inference naturally leads to either Chain-Ladder or Bornhuetter-Ferguson prediction. Using claims data on Worker’s compensation insurance from several US insurers, we discuss both frequentist and Bayesian inference.}
~
\newpage

\section{Introduction}
Claims reserving is a classical actuarial problem where actuaries estimate the outstanding liabilities of an insurer and quantify the associated variability. To emphasize its importance, first, as the largest liability item on an insurer's balance sheet, claims reserve is required to be opined by qualified actuaries to meet regulatory requirements (\cite{Friedland2013}); second, since claims for a given insurance portfolio can evolve over time, developing the incurred claims to the ultimate level is a critical component in ratemaking - another classical actuarial function for pricing the insurance contracts (\cite{BrownGottlieb2015}). In addition, reserving practice is closely related to the solvency risk. Inadequacy of reserves has been reported as the most contributing factor to a non-life insurer's failure (\cite{Coyne2008}).

In the reserving context, insurance claims data, usually referred to as ``losses", are often aggregated by lines of business and organized in a triangular format, known as ``run-off triangles", to reflect the fact that losses are incurred and developed over time. For all claims incurred in a particular year, known as the ``Accident Year", the run-off triangle shows the losses paid every year until the current calendar year. The data structure (see lower section of Table \ref{tab:triangle}) is triangular because, by the end of the evaluation year, only one year of losses would have been observed for the current accident year, while 10 years of development could have been observed for an accident year that is 10 years prior. However, it is possible that more payments relating to existing claims from an accident year can arise in the future, and also new claims corresponding to an accident year can be reported in the future. The objective is to consider the known losses so far for every accident year and obtain a forecast of incremental as well as cumulative losses for the subsequent years. The total cumulative losses resulting from any given accident year is referred to as the ``ultimate" loss. Regulatory reporting requires that such an exercise consider the recent 10 accident years, and the forecast be obtained for the 10 years, referred to as ``development years", following each accident year.

Over the years, a large variety of stochastic claims reserving methods based on run-off triangles have been proposed by practitioners and academics (see \cite{EnglandVerrall2002} for a comparison and \cite{WuthrichMerz2008} for a book-long review of alternative approaches). Among them, the most prominent and most venerable are the Chain-Ladder method and the Bornhuetter-Ferguson method. The original ideas of the Chain-Ladder and Bornhuetter-Ferguson algorithms trace back to \cite{Tarbell1934} and \cite{Bornhuetter1972}, respectively. Later, stochastic models are proposed to reproduce the prediction from the two algorithms and to quantify the associated reserving variability. For example, see the distribution-free method by \cite{Mack1993}, the bootstrap method by \cite{EnglandVerrall1999}, \cite{Peters2010}, and \cite{PinheiroAndrade2003}, and the Bayesian approach by \cite{EnglandVerrall2006} for the Chain-Ladder method; and see \cite{Verrall2004}, \cite{Mack2008} and \citeauthor{AlaiMerz2009}(\citeyear{AlaiMerz2009}, \citeyear{AlaiMerz2011}), and \cite{SaluzGisler2011} for the stochastic models that support the Bornhuetter-Ferguson method.

The Chain-Ladder and Bornhuetter-Ferguson algorithms are different yet related. Specifically, the former predicts the future cumulative losses by multiplying the current cumulative losses by suitable ``development factors" estimated from the triangle data. The latter predicts the outstanding losses by multiplying the expected ultimate losses by the percentage of (also referred to as ``quota") unpaid losses. While the percentage  of unpaid losses is estimated from the triangle data, the expected ultimate losses are usually obtained from external information, such as expert actuarial input or based on industry benchmarks. A common approach to calculate the expected ultimate loss is by taking the product of earned premiums for a given accident year and expected loss ratio (i.e., ratio of loss to premium) obtained from external sources. The link between the two algorithms is the mapping between the development factors and the development percentages (or quotas). Because of this link, the Bornhuetter-Ferguson prediction of the ultimate losses can be viewed as a credibility weighted average of the Chain-Ladder prediction based on the run-off triangle and the expected ultimate losses based on external sources.

Despite the direct relationship between the Chain-Ladder and Bornhuetter-Ferguson algorithms through the loss development pattern, there is little connection between the associated stochastic claims reserving models. Due to the need for assessing the prediction uncertainty for claims reserves, stochastic methods are independently developed to reproduce the predictions from the Chain-Ladder and Bornhuetter-Ferguson algorithms. These models are based on different assumptions and thus hardly communicate immediately to each other. As a result, the problem is typically framed in the model selection context and the selection between the two mainstream methods in practice are often ad hoc and subjective, and depending on the actuary's preference.

Motivated by the above observation, we propose a new stochastic loss reserving model based on a Dirichlet distribution (see \citealt{Frigyik2010} for an introduction to the Dirichlet distribution). The central idea is to treat the loss development quotas in a run-off triangle as compositional data and then formulate them using a Dirichlet distribution. The mathematical characterization  of the Dirichlet distribution makes it a natural choice for loss development data. \cite{darroch1971} show that a random vector $\left(x_1, \ldots, x_n, 1-\sum_{j=1}^n x_j\right)$, with each $x_i$ having a continuous probability density supported on $[0,1]$ and $0< \sum_{j=1}^{n}x_j <1$, must follow a Dirichlet distribution if for every $i$, $\frac{x_i}{1-\sum_{j\ne i} x_j}$ is independent of the vector $( x_1,\ldots, x_{i-1}, x_{i+1},\ldots, x_n)$ (i.e. all variables excluding $x_i$). In the loss development context, we can consider $x_i$ to be incremental loss in development year $i$ as a percentage of the ultimate.  The characterization amounts to the assumption that {\it the  losses known over a set of development years $(1,\ldots, i-1, i+1,\ldots, n)$ do not provide any information about the allocation of the remaining losses (i.e. $1-\sum_{j\ne i}x_j$) to the  remaining  years}. Contextually, this is a reasonable assumption. While the observed losses in a few years of development  may be suggestive of the magnitude of remaining losses per se,  they do not tell us anything about their allocation, i.e. what percentage of the remaining losses will emerge in each of  the other years. The mathematical characterization would then imply that the loss development data must follow a Dirichlet distribution. We also test this empirically by using run-off triangle data from several insurance companies.  The Dirichlet distribution is often used as a conjugate prior for the multinomial distribution in Bayesian analysis. An example of such application in loss reserving is \cite{Clark2016}. In contrast, our work, to the best of our knowledge, is the first one to employ the Dirichlet distribution for the claims data.

More importantly, the proposed Dirichlet model offers a new perspective to view the relation between two mainstream industry methods, viz. the Chain-Ladder and Bornhuetter-Ferguson methods. Interestingly, we show that the maximum likelihood estimation (MLE) of the model leads to a reserve prediction that nests the Chain-Ladder prediction. In contrast, a Bayesian inference that incorporates additional external information or expert knowledge provides the Bornhuetter-Ferguson type prediction. Therefore, the choice between the Chain-Ladder method and the Bornhuetter-Ferguson method essentially depends on the types of information available for model estimation. This is a crucial point. Because both methods can now be derived from a common stochastic reserving model, the selection of reserving methods becomes an inference problem rather than a model selection problem.

We emphasize that the proposed Dirichlet framework leads to predictions with an important desirable property in the loss reserving context. Similar to the Bornhuetter-Ferguson method, the prediction for accident-year cumulative losses is shown to be a credibility weighted average of the Chain-Ladder prediction and the expected (loss ratio) method. It is interesting that the credibility weight is determined by the coefficient of variation, as opposed to the expected value of the current cumulative losses that is used in the Bornhuetter-Ferguson method.  So, the weight assigned by the Dirichlet model not only considers the expected value but also the degree of uncertainty around the expected value. In addition, we also show that the Bornhuetter-Ferguson prediction can be obtained as special case of the Dirichlet model under some conditions.

Another contribution of the paper is that we emphasize the importance of considering a non-traditional triangle dataset to the literature and discuss its alternative usage in model inference. Specifically, we analyze run-off triangles of paid losses in workers compensation from large US property-casualty insurers. In addition to the traditional triangle data, we also have access to the claims with full development in historical accident years. We illustrate different treatment of the extra data in both frequentist and Bayesian inferences. In a case study for a particular insurer, we show that predictions based on the non-traditional dataset are better than that from the traditional dataset. Further, predictions from a Bayesian formulation that incorporate additional information turn out to be more accurate than the MLE, which does not incorporate this information. However, we suggest that the non-traditional dataset vis-a-vis traditional dataset should be used only if the additional years are believed to be representative of recent years. In general, based on a validation study conducted on 139 large insurers, we show that the Dirichlet model results are comparable to the industry benchmark, viz. the Chain-Ladder method (\citealt{Mack1993}). We find that the performance of the Dirichlet model is more consistent across accident years in its accuracy than the Chain-Ladder approach.

We also address a few methodological challenges as we implement the model. We note that the estimation and testing of the multivariate model is to be done on a special data structure, where the losses across different accident years are not identically distributed and the loss development data is incomplete for many accident years. In particular, testing goodness of fit for multivariate distributions is in general a non-trivial problem and even more so for our data structure.  We address these methodological issues.

The rest of the paper is organized as follows: Section 2 describes the data structure for the reserving setting and summarizes the sampling procedure for the study. Section 3 introduces the Dirichlet reserving model and presents the main results on reserving prediction. Section 4 discusses statistical inference for the proposed model and the method for assessing reserving variability. Section 5 performs data analysis using real run-off triangles and compares prediction with industry benchmarks. Section 6 concludes the paper.

\section{Data}

\subsection{Structure}
We consider a generic reserving setting where aggregate claims data are organized in a triangular format. The year of the incident associated with a claim is referred to as its accident year, and the subsequent years following the accident year are referred to as the development years. In a claims triangle of $m$ accident years and $n$ development years ($m>n$), we use subscripts $i(=1,\ldots,m)$ and $j(=1,\ldots,n)$ to index the accident and development years, respectively. One can interpret $i+j-1$ as the calendar year. Let $X_{ij}$ denote the incremental paid losses in accident year $i$ and development year $j$, and $E_i$ denote some known exposure that measures the volume of business in accident year $i$. Define the normalized incremental payment by $Y_{ij}=X_{ij}/E_i$. In this work, we use the earned premiums in accident year $i$ as the exposure, and we interpret $Y_{ij}$ as the loss ratio.

Table \ref{tab:triangle} visualizes the structure of the loss ratio triangle. In the table, each row shows the temporal development of losses arising from accidents occurred in a given year. Presumably claims in all accident years are settled within $n$ years, Table \ref{tab:triangle} exhibits the available claims data by the end of calendar year $m$. For the purpose of claims reserving, we are interested in the prediction of unpaid losses associated with accidents already occurred, i.e. $\{Y_{ij}: i+j-1 > m\}$. It is worth stressing that it is not necessary to assume that all claims are settled by the end of the $n$th year. This assumption is to simplify the presentation and is consistent with the real data applications. As we will show later in the text, the proposed method naturally incorporates a tail factor to allow for claims not fully developed at the largest development year.

A striking feature of Table \ref{tab:triangle} is that data are split into two sections. The lower panel of the table corresponds to a typical run-off triangle of dimension $n$ where there are $n$ rows and $n$ columns. The upper panel of the table represents additional data on historical claims that are fully developed. There are different treatments for these additional data in model inference. One could simply think of them as an additional sample for the maximum likelihood estimation. In Bayesian analysis, the data with full experience can be thought of as contributing to the updating of the prior on the unknown model parameters, which serves as an updated prior for subsequent years with incomplete experience. In addition, this data feature has an impact on model estimation specific to the proposed Dirichlet model in that the upper panel corresponds to complete observations and the lower panel corresponds to incomplete observations.

\begin{table}[htbp]
  \centering
  \caption{Exhibit of a run-off triangle of loss ratios} \label{tab:triangle}
\begin{tabular}{lclllll}
\hline\hline
&Accident & \multicolumn{5}{c}{Development Year} \\
&Year & 1 & 2 & $\cdots$ & $n-1$ & $n$  \\
\hline
&1 &$Y_{11}$ & $Y_{12}$ & $\cdots$ &  $Y_{1n-1}$ & $Y_{1n}$ \\
{\rm Fully developed}&$\vdots$&$\vdots$ &  &  & & $\vdots$  \\
&$m-n$&$Y_{m-n,1}$ & $Y_{m-n,2}$ & $\cdots$ &   $Y_{m-n,n-1}$ & $Y_{m-n,n}$ \\
\hline
&$m-n+1$&$Y_{m-n+1,1}$ & $Y_{m-n+1,2}$ & $\cdots$ &  $Y_{m-n+1,n-1}$ & $Y_{m-n+1,n}$ \\
&$m-n+2$&$Y_{m-n+2,1}$ & $Y_{m-n+2,2}$ & $\cdots$ &  $Y_{m-n+2,n-1}$ &    \\
{\rm Run-off Triangle}&$\vdots$&$\vdots$ &  & \reflectbox{$\ddots$} & &      \\
&$m-1$&$Y_{m-1,1}$ & $Y_{m-1,2}$ &  &  &    \\
&$m$&$Y_{m,1}$ &  &  &  &   \\
\hline\hline
\end{tabular}
\end{table}

\subsection{Sampling}\label{subsec:sampling}

The run-off triangle data are obtained from the Schedule P of the National Association of Insurance Commissioners (NAIC) database from years 1998-2016. The Schedule P contains firm level run-off triangles of aggregated claims
for major business lines of the U.S. property-casualty insurers. The triangles are available in terms of both incurred and paid losses. In the analysis, we examine the triangles of paid losses from the worker's compensation business. First, worker's compensation is a typical long-tailed line that demands more accurate forecast of outstanding liabilities; Second, paid losses represent realized payments and are thus less subject to measurement errors compared to incurred losses.

Data collection consists of three steps. The first step is to construct the standard run-off triangle. This portion of data is extracted from the Schedule P of year 2006. Because the Schedule P of each year contains the losses from a 10-year period up to the current calendar year, the resulting triangle includes losses that arise in accident years 1997 to 2006 and develop to year 2006 (i.e. a maximum of 10 years of development). The second step is to collect additional historical losses with full development. Since each year's Schedule P only contains losses of one accident year with 10-year development, we collect losses from additional 8 accident years, i.e. 1989-1996, from the Schedule P of years 1998-2005. The data from the first two steps form the training data that we use to develop the model. The training data of a selected insurer is illustrated in Appendix A.2. The third step is to construct the validation data, i.e. the outstanding payments to be predicted. This portion is obtained from the Schedule P of subsequent years 2007-2016. Specifically, the incremental paid losses of accident year 1998 are from the Schedule P of 2007, the incremental paid losses of accident year 1999 are from the Schedule P of 2008, and so on.

Furthermore, we restrict our analysis to large insurance groups. Specifically, we only use insurers with the minimum earned premiums from over the 18-year period greater than 100 million US dollars. This leaves us with the final 139 selected insurers. See \cite{MeyersShi2011} and \cite{Meyers2015} for more discussion on the selection of insurers for backtesting of loss reserving models.

\section{A Dirichlet Loss Reserving Model} \label{sec:model}

We propose to model the incremental loss ratios using a Dirichlet distribution. Specifically, for accident year $i$, we assume:
\begin{align} \label{equ:modeldir}
\left(\frac{Y_{i1}}{\phi_i},\cdots,\frac{Y_{in}}{\phi_i},1-\frac{\sum_{j=1}^{n} Y_{ij}}{\phi_i}\right) \sim {\rm Dir}(a_1,\ldots,a_n,b_n),
\end{align}
where $\phi_i$, $a_1,\ldots,a_n$, and $b_n$ are parameters to be estimated. For ease of notation, we denote
\begin{align}
a_0=a_1+a_2+\cdots+a_n \label{eqn:a0}.
\end{align}
 We note that ${\rm E}(\sum_{j=1}^n Y_{ij})=a_0/(a_0+b_n)\phi_i$. Thus, we can think of $\phi_i$ as the ultimate loss for accident year $i$, and $a_0/(a_0+b_n)$ as the quota of losses up to development year $n$. See Appendix A.1 for an introduction to the Dirichlet distribution and related properties. As noted in the introduction, the mathematical characterization (\citealt{darroch1971}) of the Dirichlet distribution makes it a natural choice for loss development data. According to this characterization, if we assume that {\it the  losses known over a set of development years do not give any information about how the remaining losses get allocated to the  remaining  years,} then the distribution of incremental losses (as a percentage of ultimate) must be Dirichlet.  In the loss reserving context, this is a reasonable assumption. While the observed losses in a few years of development  may be suggestive of the magnitude of remaining losses per se,  they do not inform anything about the percentage allocation of those losses to the other years.

For model (\ref{equ:modeldir}) to be legit, we require positive incremental payments and a large $\phi_i$ such that all components of the Dirichlet distribution are positive. The model does not require claims to be settled by the $n$th development year. It is easy to see that the last component in the Dirichlet model allows for tail development after $n$ years. However, one also notices that the model will require additional information to learn the tail development. Note that if the exposure data are not available, model (\ref{equ:modeldir}) certainly applies to the incremental payment triangle as well. The exposure only rescales parameter $\phi_i$. 

For a given accident year $i$, define the cumulative loss ratio from development years $k$ to $k'$, for $1\leq k \leq k'\leq n$, as $S_{i,k:k'}=\sum_{j=k}^{k'}Y_{ij}$. It is straightforward to show the following relationships:
\begin{align}
{\rm E}(S_{i,k:k'})&=\frac{\sum_{j=k}^{k'}a_j}{a_0+b_n}\phi_i, \label{equ:clmean}\\
{\rm Var}(S_{i,k:k'})&=\frac{\left(\sum_{j=k}^{k'}a_j\right)\left(a_0+b_n-\sum_{j=k}^{k'}a_j\right)}{(a_0+b_n)^2(a_0+b_n+1)}\phi_i^2.\label{equ:clvar}
\end{align}
This provides interesting interpretations for the model parameters. Considering the case $k=k'$, one could interpret $a_j/(a_0+b_n)$ as the development percentage in development year $j$, and $\phi_i$ as the expected ultimate loss ratio in accident year $i$.

At any development year $k$, model (\ref{equ:modeldir}) further implies the following about the conditional distributions given the cumulative loss ratio $S_{i,1:k}$:
\begin{align}
\left(\frac{Y_{i1}}{S_{i,1:k}},\cdots,\frac{Y_{ik}}{S_{i,1:k}}\right) &|S_{i,1:k} \sim {\rm Dir}(a_1,\ldots,a_k), \label{equ:cond1}\\
\left(\frac{Y_{ik+1}}{\phi_i-S_{i,1:k}},\cdots,\frac{Y_{in}}{\phi_i-S_{i,1:k}},\frac{\phi_i-S_{i,1:n}}{\phi_i-S_{i,1:k}}\right) &|S_{i,1:k} \sim {\rm Dir}(a_{k+1},\ldots,a_n,b_n). \label{equ:cond2}
\end{align}
The above relations also yield intuitive interpretations. Suppose the evaluation year is $i+k-1$, then equation (\ref{equ:cond2}) provides an update on the future development pattern. Specifically, one could interpret $a_j/(\sum_{j=k+1}^n a_j+b_n)$ ($j\ge k+1$) as the disposal rate for development year $j$ (see \cite{BrownGottlieb2015} for the closure method in loss reserving).

In the context of claims reserving, one outcome of particular interest is the total outstanding liability of the insurer. According to the Dirichlet model, the unpaid losses at the end of development year $k$ for accident year $i$, $S_{i,k+1:n}$, follows a scaled Beta distribution as follows:
\begin{align}
\frac{S_{i,k+1:n}}{\phi_i-S_{i,1:k}}|S_{i,1:k} \sim {\rm Beta}\left(a_0-\sum_{j=1}^k a_j, b_n\right).
\end{align}

To establish connection between the predictions from the proposed reserving model and industry benchmarks, we define the loss development factor and the loss development quota following \cite{SchmidtZocher2016}. Specifically, define $\gamma_{k:k+1}$ the loss development factor from development year $k$ to development year $k + 1$, and $\eta_{k}$ the development quota for the $k$th development year over a $n$-year period as:
\begin{align} \label{equ:factorquota}
\gamma_{k:k+1} = \frac{{\rm E}(S_{i,1:k+1})}{{\rm E}(S_{i,1:k})}, \quad {\rm and} \quad \eta_{k} = \frac{{\rm E}(S_{i,1:k})}{{\rm E}(S_{i,1:n})}.
\end{align}
Furthermore, the development factors and development quotas satisfy the following relationship:
\begin{align*}
\gamma_{k:k+1} = \frac{\eta_{k+1}}{\eta_{k}}, \quad {\rm or} \quad \eta_{k} = \prod_{j=k}^{n-1} \frac{1}{\gamma_{j:j+1}}.
\end{align*}
Under the proposed Dirichlet model, we have:
\begin{align}
\gamma_{k:k+1} &=\frac{a_1+\cdots+a_{k+1}}{a_1+\cdots+a_{k}}, \label{equ:devfactor}\\
\eta_{k} &= \frac{a_1+\cdots+a_{k}}{a_1+\cdots+a_{n}}. \label{equ:devquota}
\end{align}

Below we use $\widehat{Y}$ and $\widehat{S}$ to denote the prediction for the incremental and cumulative payments respectively.
The loss reserve at the end of the $k$th evaluation year for accident year $i$, is defined as the predicted total unpaid loss for the $i$th accident year from development years $k+1$ to $n$. We denote the loss reserve from different methods by $\widehat{R}_{i}$ with the appropriate superscript:
\begin{align}\label{equ:reserve}
\widehat{R}_{i} = \widehat{S}_{i, 1:n}- S_{i, 1:k}.
\end{align}
To facilitate comparison,  we use superscript ``D'', ``CL'', ``EX'', and ``BF'' to denote the Dirichlet method, the Chain-Ladder or development method, the expected (loss ratio) method, and the Bornhuetter-Ferguson method, respectively.

\subsection{Predictions from Industry Benchmarks}
The Chain-Ladder method and the Bornhuetter-Ferguson method are the two golden benchmarks widely used by practitioners for setting loss reserves for property and casualty business lines. The Chain-Ladder method assumes that the expected cumulative losses up to development year $k+1$, conditional on the paid losses up to age $k$, can be obtained as a factor multiple of losses up to age $k$, i.e.,
 \begin{eqnarray}
{\rm E}(S_{i,1:k+1}\vert S_{i,1:k})= \gamma_{k:k+1} S_{i,1:k} \quad {\rm for} \quad k=1,\ldots,n-1. \label{equ:CLexpcumlossk}
 \end{eqnarray}
It follows that the Chain-Ladder predictions, at the end of the $k$th evaluation year, for the incremental paid losses in development year $k'$($>k$) and the cumulative paid losses at the end of development year $n$ are:
  \begin{align}
\widehat{Y}_{ik'}^{CL} &= {\rm E}(Y_{i,k'}\vert S_{i,1:k})= S_{i,1:k} \left(\prod_{j=k}^{k'-1}{\gamma}_{j:j+1}-\prod_{j=k}^{k'-2}{\gamma}_{j:j+1}\right), \label{equ:CLpredYton} \\
\widehat{S}^{CL}_{1:n} &= {\rm E}(S_{i,1:n}\vert S_{i,1:k})= \prod_{j=k}^{n-1} {\gamma}_{j:j+1}~S_{i,1:k}= \dfrac{1}{{\eta}_k}~S_{i,1:k}. \label{equ:CLExpcumlossn}  %
 \end{align}
The validity of the Chain-Ladder prediction relies on the stable operation of the insurer. If there is some underlying change in the insurance operation such as underwriting criterion or settlement practice, the approach could lead to unreasonable predictions.

The Bornhuetter-Ferguson method addresses the above issue by assuming that the expected cumulative paid losses is a fixed portion of the ultimate losses which could be obtained using either the internal or external information. Specifically, the method assumes:
 \begin{eqnarray}
{\rm E}(S_{i,1:k})= \eta_k~ \widehat{S}_{i,1:n}^{EX} \quad {\rm for} \quad k=1,\ldots,n, \label{equ:BFexpcumlossk}
 \end{eqnarray}
where $\eta_k$ is interpreted as the percentage of cumulative paid losses by development year $k$, and $\widehat{S}_{i,1:n}^{EX}$ represents the expected cumulative loss ratio over $n$ development years. When $\widehat{S}_{i,1:n}^{EX}$ is determined using external data, the Bornhuetter-Ferguson prediction is less subject to the distortion caused by the operational change. Under this assumption, the prediction for the incremental, cumulative paid losses and reserves are shown as:
\begin{align}
\widehat{Y}_{ik'}^{BF} &= {\rm E}(Y_{i,k'}\vert S_{i,1:k})= ({\eta}_k-{\eta}_{k-1}) \widehat{S}_{i,1:n}^{EX},\\
\widehat{S}_{i, 1:n}^{BF} &= {\rm E}(S_{i,1:n}\vert S_{i,1:k})= {\eta}_k \widehat{S}_{i, 1:n}^{CL} + (1-{\eta}_k)\widehat{S}_{i, 1:n}^{EX},\\
\widehat{R}_{i}^{BF} &= {\rm E}(S_{i,1:n}\vert S_{i,1:k})-S_{i,1:k}= {\eta}_k \widehat{R}_{i}^{CL} + (1-{\eta}_k)\widehat{R}_{i}^{EX}. \label{eqn:BFreserve}
\end{align}
The Bornhuetter-Ferguson prediction can be interpreted as a weighted average of the Chain-Ladder prediction and the expected method prediction. The result can be easily shown using the relationship between the development factor and the development quota. Note that the above forecasts are predictors that represent the theoretical expected payments from different development assumptions. To quantify the reserving variability, one needs to take into account the process variance and the parameter uncertainty.
\subsection{Predictions from Dirichlet Model}
This section summarizes our main result on the loss reserving prediction using the proposed Dirichlet method for run-off triangles. Similar to Bornhuetter-Ferguson, the prediction from the Dirichlet model results in a weighted average of the Chain-Ladder prediction and the expected method prediction, but with the credibility weights that consider the degree of  uncertainty in addition to the expected value.

\bigskip
\textit{Proposition 1.} If incremental paid losses $\{Y_{ij}: i=1,\ldots,m; j=1\ldots,n\}$ follow the Dirichlet model (\ref{equ:modeldir}), the cumulative losses as well as the loss reserves for the $i$th accident year at the end of the $k$th evaluation year can be expressed as a weighted average of the development method and expected method. The weight is determined by the coefficient of variation of cumulative payments. To be more specific,
\begin{align}
 \widehat{S}^{D}_{i,1:n} &=v(k) \widehat{S}^{CL}_{1:n}+ (1-v(k)) \widehat{S}^{EX}_{1:n}, \label{equ:result2cum}\\
\widehat{R}_{i}^{D} &= {v}(k) \widehat{R}_{i}^{CL} + (1-{v}(k))\widehat{R}_{i}^{EX} \label{equ:result2},
\end{align}
with
\begin{align*}
v(k) &= \left\{\frac{{\rm CV}(S_{i,1:n})}{{\rm CV}(S_{i,1:k})}\right\}^2 = \frac{{\rm Var}(S_{i,1:n})}{{\rm Var}(S_{i,1:k})}\left\{\frac{{\rm E}(S_{i,1:k})}{{\rm E}(S_{i,1:n})}\right\}^{2},\\
\widehat{R}_{i}^{CL} & = S_{i,1:k} \left(\prod_{j=k}^{n-1}\gamma_{j:j+1}-1\right), \\
\widehat{R}_{i}^{EX} & = {\rm E}(S_{i,1:n})-S_{i,1:k}.
\end{align*}

Proof. At the $k$th evaluation year for accident year $i$, cumulative losses upto $k$ development years ($S_{i, 1:k}$) is known. So, the prediction from the Dirichlet model $\widehat{S}^{D}_{i,1:n}$ is obtained as:
\begin{align*}
&{\rm E}(S_{i,1:n}|S_{i,1:k})\\
=&S_{i,1:k}+\frac{\sum_{j=k+1}^{n}a_j}{\sum_{j=k+1}^{n}a_j+b_n}(\phi_i-S_{i,1:k}) \\
=&\frac{\sum_{j=n+1}^{n}a_j+b_n}{\sum_{j=k+1}^{n}a_j+b_n}\frac{\sum_{j=1}^{k}a_j}{\sum_{j=1}^{n}a_j}\left(\frac{\sum_{j=1}^{n}a_j}{\sum_{j=1}^{k}a_j}S_{i,1:k}\right) +\frac{\sum_{j=k+1}^{n}a_j}{\sum_{j=k+1}^{n}a_j+b_n}\frac{\sum_{j=1}^{n}a_j+b_n}{\sum_{j=1}^{n}a_j}\left(\frac{\sum_{j=1}^{n}a_j}{\sum_{j=1}^{n}a_j+b_n}\phi_i\right)\\
=&\left(\frac{b_n}{\sum_{j=k+1}^{n}a_j+b_n}\frac{\sum_{j=1}^{k}a_j}{\sum_{j=1}^{n}a_j}\right)  S_{i,1:k} \prod_{j=k}^{n-1}\gamma_{j:j+1} +\left(1-\frac{b_n}{\sum_{j=k+1}^{n}a_j+b_n}\frac{\sum_{j=1}^{k}a_j}{\sum_{j=1}^{n}a_j} \right) {\rm E}(S_{i,1:n})\\
=& v(k) \widehat{S}^{CL}_{1:n}+ (1-v(k)) \widehat{S}^{EX}_{1:n}.
\end{align*}
Equation (\ref{equ:result2cum}) follows because one can show using equations  (\ref{equ:clmean}) and (\ref{equ:clvar}) that
\begin{align}\label{equ:CLweight}
v(k)=\frac{b_n}{\sum_{j=k+1}^{n}a_j+b_n}\frac{\sum_{j=1}^{k}a_j}{\sum_{j=1}^{n}a_j} = \left\{\frac{{\rm CV}(S_{i,1:n})}{{\rm CV}(S_{i,1:k})}\right\}^2.
\end{align}
Equation (\ref{equ:result2}) is obtained based on $\widehat{R}_{i}^{D}= E(S_{i,1:n}|S_{i,1:k}) - S_{i,1:k}$. \quad $\Box$

\bigskip
\textit{Remark.} To compare the Dirichlet method with the Bornhuetter-Ferguson method, recall
\begin{align*}
\widehat{R}_{i}^{BF} = {\eta}_k \widehat{R}_{i}^{CL} + (1-{\eta}_k)\widehat{R}_{i}^{EX} \mbox{ (see equation (\ref{eqn:BFreserve})) },
\end{align*}
where $\eta_k$ is defined in equation (\ref{equ:factorquota}). Both the Dirichlet method and the Bornhuetter-Ferguson method express reserves as a weighted average of the predictions from the Chain-Ladder and expected methods. The difference lies in the credibility weight given to the Chain-Ladder prediction. Under the proposed model (\ref{equ:modeldir}), the Bornhuetter-Ferguson method gives higher weight when the cumulative loss ratio at the time of valuation has higher expected value, while the Dirichlet method assigns higher weight when the cumulative loss ratio at the time of valuation has lower coefficient of variation. So, the weight assigned by the Dirichlet model not only considers the expected value but also the degree of uncertainty around the expected value. Further from (\ref{equ:devquota}) one notes $v(k)\approx \eta_k$ when $\frac{\sum_{j=k+1}^{n}a_j}{b_n}\approx 0$, i.e. when the expected loss development from year $(k+1)$ to year $n$ is negligible compared to the tail development after the $n$th year, the Dirichlet weight reduces to the Bornhuetter-Ferguson weight. Heuristically, if the development from $k+1$ up to $n$ is negligible, then there is more certainty around the development up to $n$ and hence variance does not play a role.
\section{Statistical Inference}  \label{sec:inference}
This section focuses on the statistical inference for the proposed Dirichlet model. We show that two disparate but widely used industry approaches, viz. Chain-Ladder and Bornhuetter-Ferguson, naturally result from the proposed Dirichlet model, thus providing a common statistical framework for the approaches. The choice between the two approaches is then driven by the nature of information supplied to the model, rather than a subjective choice made by the analyst.

 We present two alternative strategies, maximum likelihood estimation and Bayesian method, based on a realized sample of Table \ref{tab:triangle}. We note the strength and limitations for each strategy and make recommendations regarding when each approach is suitable. In the reserving context, actuaries are interested in an interval prediction which leads to more informative decision making. To quantify reserving variability, one has to account for both process uncertainty and parameter uncertainty. For this reason, we also discuss, for each inference method, the general steps to obtain the predictive distribution for claims reserves. In the following, we use $\bm{\theta}$ to denote the vector of all model parameters, i.e.
\begin{align} \label{equ:modelpar}
\bm{\theta}=\left(a_1, a_2, \ldots, a_n, b_n, \phi_1, \phi_2, \ldots, \phi_m \right).
\end{align}

\subsection{Maximum Likelihood Estimation} \label{subsec:mle}
Here, we describe the maximum likelihood estimation for  model \ref{equ:modeldir} based on data in Table \ref{tab:triangle}. We note that:

for $1\leq i \leq m-n$,
\begin{align*}
\left(\frac{Y_{i1}}{\phi_i},\cdots,\frac{Y_{in}}{\phi_i},1-\frac{S_{i,1:n}}{\phi_i}\right) \sim {\rm Dir}(a_1,\ldots,a_n,b_n),
\end{align*}

for $m-n+1\leq i \leq m$,
\begin{align*}
\left(\frac{Y_{i1}}{\phi_i},\cdots,\frac{Y_{im+1-i}}{\phi_i},1-\frac{S_{i,1:m+1-i}}{\phi_i}\right) \sim {\rm Dir}\left(a_1,\ldots,a_n,a_0+b_n-\sum_{j=1}^{m+1-i}a_j\right).
\end{align*}
Thus the likelihood function for the $i$th accident year is:
\begin{align}\label{equ:li}
l_i(\bm{\theta}) = \left\{
        \begin{array}{ll}
          \cfrac{\Gamma(a_0+b_n)}{\prod\limits_{j=1}^{n}\Gamma(a_j)\Gamma(b_n)}\left(\cfrac{1}{\phi_i}\right)^{n}\prod\limits_{j=1}^{n}\left(\cfrac{y_{ij}}{\phi_i}\right)^{a_j-1}\left(1-\cfrac{s_{i,1:n}}{\phi_i}\right)^{b_n-1}, & 1\leq i \leq m-n \\
          \cfrac{\Gamma(a_0+b_n)}{\prod\limits_{j=1}^{m+1-i}\Gamma(a_j)\Gamma\left(a_0+b_n-\sum\limits_{j=1}^{m+1-i}a_j\right)}\left(\cfrac{1}{\phi_i}\right)^{m+1-i}\prod\limits_{j=1}^{m+1-i}\left(\cfrac{y_{ij}}{\phi_i}\right)^{a_j-1} &  \\
          \quad\quad\quad\quad \times \left(1-\cfrac{s_{i,1:m+1-i}}{\phi_i}\right)^{a_0+b_n-\sum\limits_{j=1}^{m+1-i}a_j-1}, & m-n+1\leq i \leq m \\
        \end{array}.
      \right.
\end{align}
where $y$ and $s$ are realized values of incremental and cumulative paid loss ratios respectively. Define the total loglikelihood function as
\begin{align}
ll(\bm{\theta})=\sum_{i=1}^{m}\ln l_i(\bm{\theta}).\label{eqn:logl}
\end{align}
 The maximum likelihood estimator of $\bm{\theta}$ is
\begin{align}
\widehat{\bm{\theta}}^{MLE} = \argmax_{\bm{\theta}} ll(\bm{\theta}).\label{eqn:maxlik}
\end{align}
For meaningful estimation using MLE, we note that it is necessary to have the condition $b_n\geq 1$. If $b_n<1$, then the likelihood can be made infinity by choosing $\phi_i=s_{i,1:n}$ for $1\leq i\leq m-n$ or $\phi_i=s_{i,m+1-i}$ for $m-n+1\leq i\leq m$. Assuming $b_n\geq 1$  essentially amounts to assuming that the distribution of cumulative losses $S_{i,1:n}$ is uni-modal. The condition $b_n\geq 1$ also ensures that the MLE for $\phi_i$ is necessarily greater than the observed cumulative losses in the data for the accident year $i$. Detailed derivation of MLE is given in the Appendix A.3. The MLE procedure is coded and implemented using $R$ \cite{rcore2013}. We note that the MLE of $b_n$ is given by $\widehat{b}_n=1$, and the MLE of $\phi_i$ is
\begin{align} \label{equ:mlephi2}
\widehat{\phi}_i = \left\{
           \begin{array}{ll}
             s_{i,1:n}, & 1\leq i \leq m-n \\
             \cfrac{\widehat{a}_0}{\sum_{j=1}^{m+1-i}\widehat{a}_j}s_{i,1:m+1-i}, & m-n+1\leq i \leq m .
           \end{array}
         \right.
\end{align}
In absence of additional information on $\phi_i$ or other parameters, the MLE of the Dirichlet model leads to a reserve prediction that nests the Chain-Ladder prediction. To see this, recall from \textit{Proposition 1} that the prediction of cumulative losses in accident year $i ~(m-n+1 < i \leq m)$ is:
\begin{align*}
\widehat{S}^{D}_{i,1:n} &={\rm E}(S_{i,n}|S_{i,m+1-i}=s_{i,m+1-i})\\
&= \left\{{v}(m+1-i) + (1-{v}(m+1-i))\frac{{a}_0}{{a}_0+1}\right\}s_{i,1:m+1-i}\frac{{a}_0}{\sum_{j=1}^{m+1-i}{a}_j}\\
&\overset{{a}_0/({a}_0+1)\rightarrow 1}{\approx} s_{i,1:m+1-i}\prod_{j=m+1-i}^{n-1}{\gamma}_{j:j+1}~~=~~ \widehat{S}^{CL}_{i,1:n}.
\end{align*}
Thus the Chain-Ladder prediction is obtained as a limiting case of the Dirichlet model using MLE for inference. We verify in Section \ref{sec:dataanalysis} that the condition ${a}_0/({a}_0+1)\approx 1$ is supported by the real run-off triangle data in the empirical study.

To quantify the reserving uncertainty, one could resort to the parametric bootstrap method. Specifically, parametric bootstrap requires the steps below to obtain the predictive distribution of unpaid losses $\{Y_{ij}: m-n+1 < i\leq m, m+1 \le i+j \leq m+n\}$:
\begin{itemize}
\item[1)] Given $\widehat{\bm{\theta}}^{MLE}$, generate data of paid losses $\mathcal{D}_{U}^{(s)} = \{y_{ij}^{(s)}: 1 \leq j\leq n,  i+j \leq m+1\}$ from model (\ref{equ:modeldir});
\item[2)] Use data $\mathcal{D}_{U}^{(s)}$ to estimate $\bm{\theta}$, denoting the estimates as $\widehat{\bm{\theta}}^{(s)}$;
\item[3)] Given $\widehat{\bm{\theta}}^{(s)}$, simulate data of unpaid losses $\mathcal{D}_{L}^{(s)} = \{y^{(s)}_{ij}: m-n+1 < i\leq m, 1\leq j \leq n, m+1 < i+j \}$ from model (\ref{equ:cond2});
\item[4)] Repeat steps 1)-3) for $s=1,\ldots,n_{sim}$, where $n_{sim}$ is the number of bootstrap samples. One obtains the distribution of $\widehat{\bm{\theta}}^{MLE}$ and the predictive distribution of unpaid losses $\mathcal{D}_{L}$.
\end{itemize}
However, it is the characteristic of problems where one of the parameters is on the boundary of the support of the distribution, that MLE can be biased and bootstrap can lead to  biased sampling of the parameters (see e.g. \citealt{andrews2000}, \citealt{hall2002}). For our scaled Dirichlet model, the scale parameter $\phi_i$ happens to be the upper end point for the support of the distribution of $S_{i, 1:n}$. To our knowledge, we are not aware any methods for bias correction of bootstrap parameters for the scaled Dirichlet model. Here, we propose and implement a computational approach, based on a two-stage bootstrap procedure to correct for the bias in the bootstrap samples. The details of the approach are summarized in Appendix A.4.
\subsection{Testing Goodness of Fit}\label{subsec:GOF}
Loss reserving is an exercise carried out for an individual insurance company. Hence, it is of interest to have a method to formally test whether the proposed Dirichlet model is a good fit for a given insurance company, based on the available loss data. Goodness of fit test for multivariate distributions is in general less straight forward.  Tests specific to multivariate normal distribution are more commonly studied (e.g. \citealt{gallardo1979}, \citealt{paulson1987}, \citealt{surucu2006}). Notable distribution-free approaches include an extension of Kolmogorov-Smirnov test by \cite{juste1997} and  an extension of Cramer-von-Mises test by \cite{chiu2009}. The distribution-free approaches are based on iid data and require the construction of the empirical distribution function. A recent thesis by \cite{Li2015} considers tests for the Dirichlet distribution, but is also based on iid and complete data. It does not appear easy to extend these approaches to testing the scaled Dirichlet model where the loss data are non-iid and incomplete.   However, partly motivated by ideas in the afore-mentioned works, we devise an approach to test whether the Dirichlet model is a reasonable fit to the observed loss data for a given insurance company.  Denote the observed data for any given company by ${\bf y}_{obs}$. We want to test
\[ H_0: {\bf y}_{obs}\sim\mbox{Dirichlet model (\ref{equ:modeldir})} \mbox{ vs. } H_1: ~{\bf y}_{obs} \mbox{ does not follow } (\ref{equ:modeldir}). \]
Our approach to testing is based on the property that marginals of the Dirichlet distribution are also Dirichlet, and its relation to the Beta distribution. If the null hypothesis is true, then for any $k<n$, and for any $i=1,2,\ldots, m$, we have
\[\left(\frac{Y_{i1}}{\phi_i},\frac{Y_{i2}}{\phi_i}, \ldots, \frac{Y_{ik}}{\phi_i} \right) \sim Dir\left(a_1,a_2,\ldots, a_k, \sum_{j=k+1}^n a_j +b_n\right).\]
It can be shown that this holds if and only if
\[\begin{cases} \frac{Y_{i1}}{\phi_i} \sim Beta \left(a_1, a_2+\ldots+a_n+b_n\right) \mbox{ and} \\  \frac{Y_{i2}}{\phi_i-Y_{i1}} \sim Beta\left(a_2, a_3+\ldots+a_n+b_n\right) \mbox{ and} \\
~~~\vdots \\
\frac{Y_{ik}}{\phi_i-Y_{i1}-\ldots-Y_{i (k-1)}} \sim Beta(a_k, a_{k+1}+\ldots+a_n+b_n). \end{cases}\]
Therefore, for any accident year $i$ with observed losses for $k$ development years, if we denote the cdfs of the Beta distribution for $\frac{Y_{i1}}{\phi_i}, \frac{Y_{i2}}{\phi_i-Y_{i1}} , \ldots, \frac{Y_{ik}}{\phi_i-Y_{i1}-\ldots-Y_{i (k-1)}}  $, by $F_{i1}, F_{i2}, \ldots, F_{ik}$ respectively, then $\forall ~i\in \{1,2,\ldots, m\}, k\in \{1,2,\ldots, \min(n, m-i+1)\}$,
\[F_{i1}\left(\frac{Y_{i1}}{\phi_i}\right), F_{i2}\left(\frac{Y_{i2}}{\phi_i-Y_{i1}}\right), \ldots, F_{ik}\left(\frac{Y_{ik}}{\phi_i-Y_{i1}-\ldots-Y_{i (k-1)}}\right) \stackrel{iid}{\sim} uniform(0,1).\]
We test for the above based on the Kolmogorov-Smirnov statistic, which we denote by $T({\bf y})$ and is computed using the data
\[ \left\{ \widehat{F}_{ik}\left(\frac{Y_{ik}}{\phi_i-\sum_{j=1}^{k-1}Y_{ij}}\right)~, \mbox{ for } (i,k) \in ~\mathcal{S}  \right\}, \]
where,
\[\mathcal{S}=\left\{(i,k): i\in\{1,2,\ldots,m\},j\in\{1,2,\ldots,\min(n,m-i+1)\right\}\backslash \left\{(i,n): i=1,2,\ldots, m-n \right\}. \]
As a practical matter, we note that the computation of estimated cdfs above are based on the respective Beta distributions using the MLEs of the parameters, based on the given data ${\bf y}$. We note  using equation (\ref{equ:mlephi}), that the MLE for $\widehat{\phi}_i =s_{i, 1:n}$ whenever $i\leq m-n$, which leads to degeneracy in the distribution, since $1-\frac{\sum_{j=1}^{n}Y_{ij}}{\widehat{\phi}_i}=0$. So, for $i\leq m-n$, we exclude $k=n$.

Our testing procedure based on the test-statistic $T({\bf y})$ is as follows
\begin{itemize}
\item[] {1.} Determine the distribution of $T({\bf y})$ under the null-hypothesis by:
\item[] ~~~ (a) Draw several bootstrap samples ${\bf y}$ using Dirichlet model (\ref{eqn:logl}), with $\bm{\theta}=\widehat{\bm{\theta}}^{MLE}({\bf y}_{obs})$.
\item[] ~~~{(b)} For each draw we compute  $T({\bf y})$, and finally obtain their empirical distribution.
\item[] {2.} At significance level $\alpha$ (for instance, 5\%), compute the $100(1-\alpha)\%$ confidence region by marking the $\alpha/2$ and $(1-\alpha/2)$ quantile of the empiricial null distribution of the test-statistic.
\item[] {3.} If $T({\bf y}_{obs})$ does not belong to the $100(1-\alpha)\%$ confidence region, reject $H_0$ at $\alpha$ significance level.
\end{itemize}

\subsection{Bayesian Method} \label{subsec:bayes}
Bayesian method enjoys a couple of advantages in the claims reserving applications. First, it allows one to incorporate external information which could be expert knowledge or additional data into model inference (see, for example, \cite{ZhangDukic2012}, \cite{ShiHartman2016}, and \cite{Shi2017}). Second, it integrates the estimation and prediction and thus makes it straightforward to quantify the reserving uncertainty. Recall that we use $\mathcal{D}_U$ and $\mathcal{D}_L$ to denote the observed paid losses and the unpaid losses to be predicted respectively. Let $p(\bm{\theta})$ and $p(\bm{\theta}|\mathcal{D}_U)$ denote the prior and posterior distributions for parameters $\bm{\theta}$. The general idea to obtain the predictive distribution of unpaid losses is (see \cite{Gelman2004}):
\begin{align*}
f(\mathcal{D}_L|\mathcal{D}_U)=\int f(\mathcal{D}_U|\bm{\theta})p(\bm{\theta}|\mathcal{D}_U) d\bm{\theta},
\end{align*}
where $p(\bm{\theta}|\mathcal{D}_U)\propto f(\mathcal{D}_U|\bm{\theta})p(\bm{\theta})$ and $ f(\mathcal{D}_U|\bm{\theta})=\prod_{i=1}^m l_i(\bm{\theta})$, with $l_i(\bm{\theta})$ as in equation (\ref{equ:li}).

In particular, the Bayesian inference offers extra flexibility for the proposed Dirichlet model in several ways. First, using the additional years of data with fully developed losses, along with the recent years where losses are not fully developed,  contributes to our knowledge about the unknown parameters, when we use the posterior distribution for predicting the losses for recent years.  Second, the Bayesian method allows for a natural hierarchical extension of the Dirichlet model. This would be sensible when one can reasonably assume that the operations of the company are identical across accident years, and that the resulting variations in the ultimate loss ratios are only due to random variations in the operations across years. Third, expert knowledge on tail factors could be intuitively integrated into the inference. Recall that ${\rm E}(S_{i,1:n})=a_0/(a_0+b_n)\phi_i$. Thus, it is intuitive to think of $\phi_i$ as the ultimate losses and interpret $1+b_n/a_0$ as the tail factor. Interestingly, we have already seen that the MLE of $\phi_i$ in equation (\ref{equ:mlephi2}) is consistent with a tail factor of one because it does not use any external information in the estimation. With additional knowledge on the tail factor, one could impose an informative prior on $b_n/a_0$ in the inference.  Thus the prior could incorporate knowledge from both internal and external data in this context. In Section \ref{sec:bayesinference}, we consider the hierarchical prior specification $\phi_1, \phi_2, \ldots, \phi_n \stackrel{iid}{\sim} ~ uniform(0, \phi)$ with a hyper prior $p(\phi)\propto 1$. We also consider a flat prior for $b_n$ with appropriate conditions on its support to account for the knowledge on the tail development of claims.
\section{Data Analysis}\label{sec:dataanalysis}

In the application, we examine the run-off triangle data of worker's compensation paid losses from US property-casualty insurers. The data of each individual company contain incremental paid losses for 18 accident years ($m=18$) from to 1989 to 2006, and for each accident year, losses are developed for the period of 10 years ($n=10$). In addition, the earned premiums are available for each accident year. We calculate the incremental loss ratios by dividing the paid losses by the earned premiums. Then we split the data into two segments, the upper triangle data $\mathcal{D}_{U}=\{y_{ij}: 1\leq j \leq 10, i+j\leq 19 \}$ and the lower triangle data
$\mathcal{D}_{L}=\{y_{ij}: 10\leq i \leq 18, 1\leq j \leq 10, 20 \leq i+j\}$. We use the upper triangle data $\mathcal{D}_{U}$ to develop the model and the lower triangle data $\mathcal{D}_{U}$ to validate the prediction.

\begin{figure}
\caption{\label{fig:tsplot} Time series plot of cumulative loss ratios by accident year for the selected insurer. }
    \centering
    \subfigure[Fully developed claims]{\label{fig:a}\includegraphics[width=70mm]{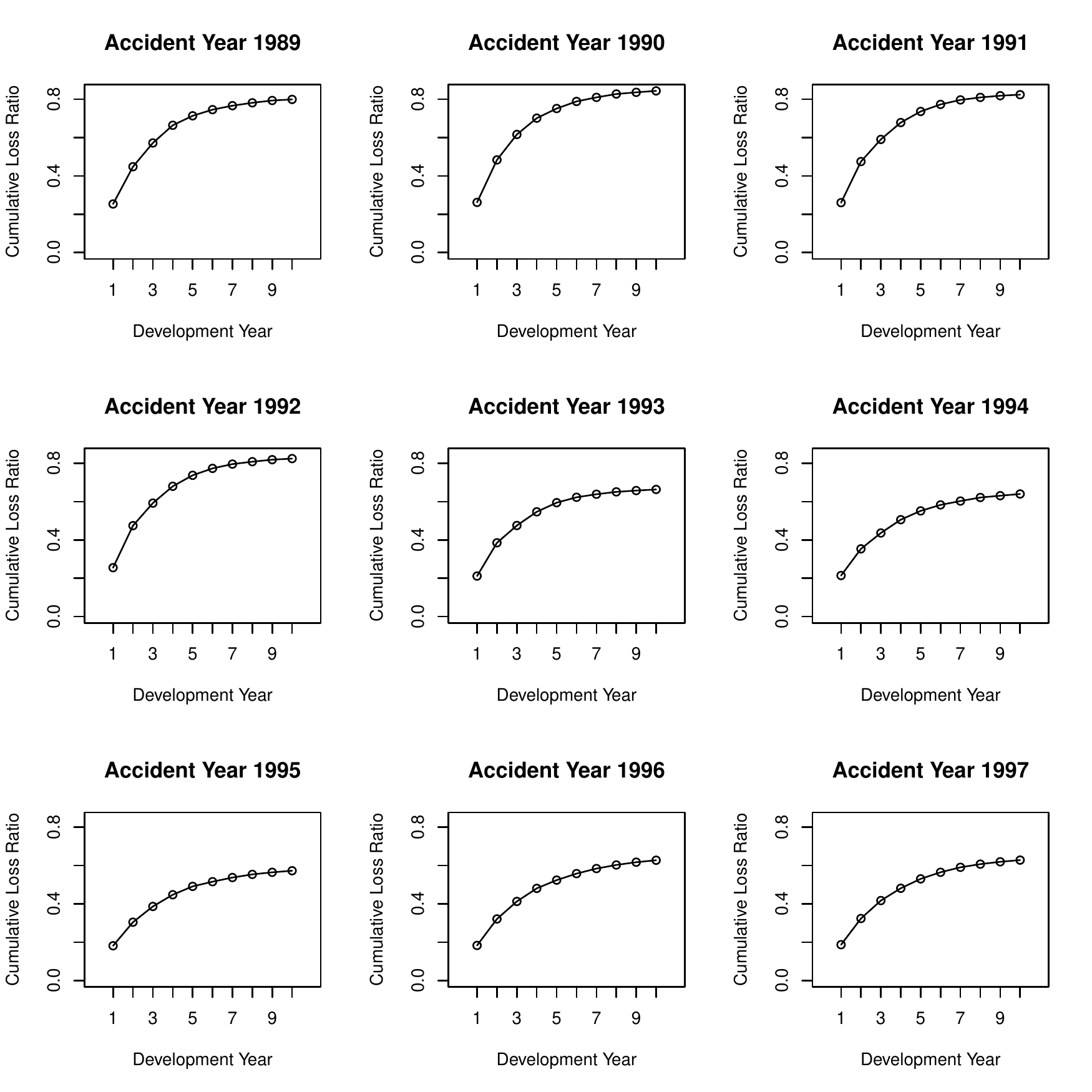}}
\subfigure[Partially developed claims]{\label{fig:b}\includegraphics[width=70mm]{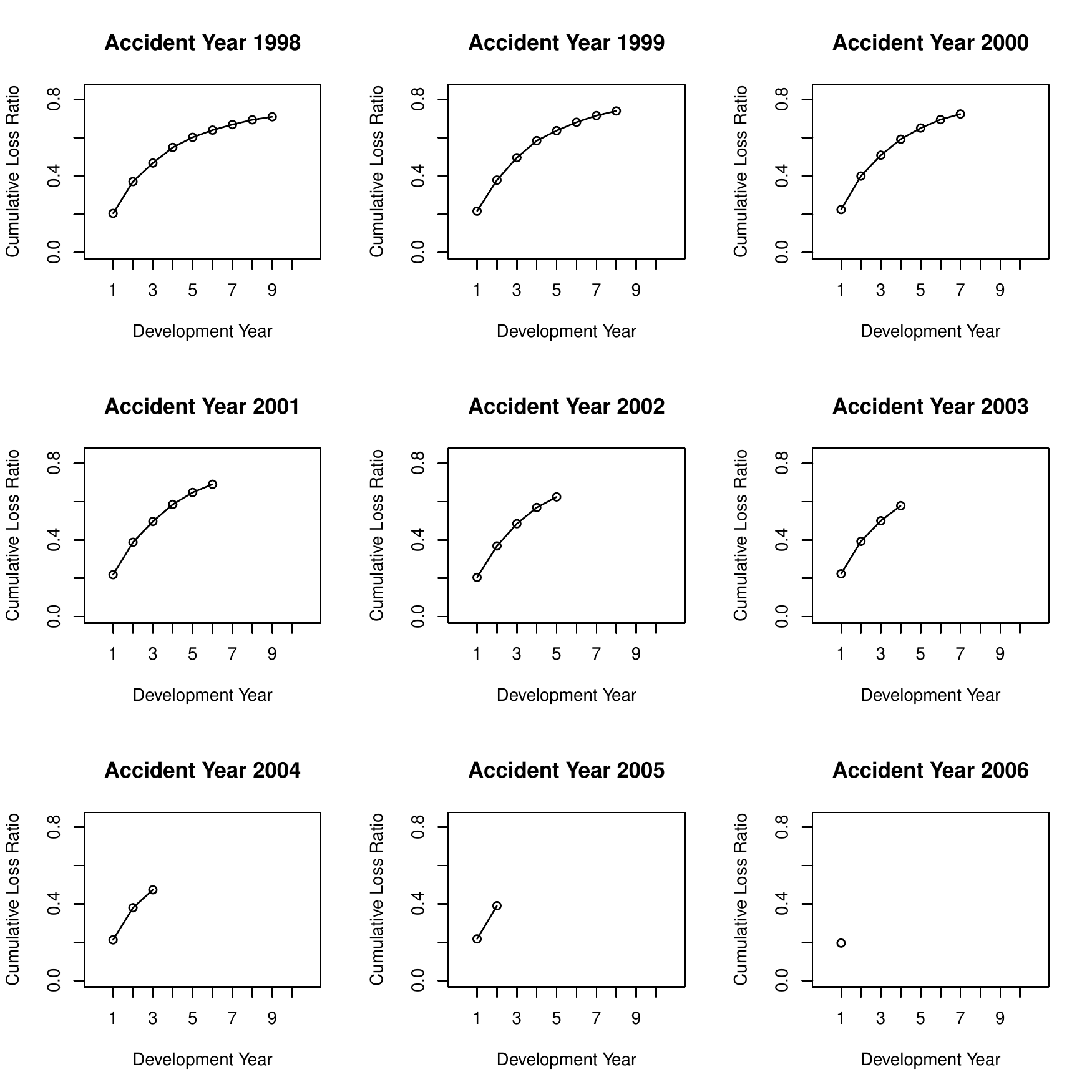}}
\end{figure}

In this section, we demonstrate the application of the model for one selected insurer and also summarize the performance of our approach by applying it to 139 large insurers in the NAIC database selected based on criteria described in Section \ref{subsec:sampling}.  To visualize the data, we first select one large insurer and exhibit in Figure \ref{fig:tsplot} the time series plot of loss ratios by accident year. The first panel shows the loss development for the first 9 accident years, where the claims are fully developed up to 10 development years, and the second panel for the last 9 accident years, where the development is incomplete, i.e. it is not yet fully known. The accident years 1989 to 1996 correspond to the additional information that one could use for the estimation of the model, and the accident years 1997 to 2006 represent the typical run-off triangle data. The raw premium and loss data are provided in Appendix A.2.

\subsection{Estimation} \label{subsec:est}
Table \ref{T:Dparest} shows the maximum likelihood estimates and the associated standard errors (SE) for the Dirichlet model (\ref{equ:modeldir}) based on data from 10 accident years of data (panel (b) in Figure \ref{fig:tsplot}), as well as from 18 accident years (panels (a) and (b) in Figure \ref{fig:tsplot}). The standard errors are obtained using the bootstrap method described in Section \ref{subsec:mle}. We note here that for this selected insurer,  the goodness-of-fit test described in Section \ref{subsec:GOF}  supports the Dirichlet-model assumption (at 5\% level). This holds true when we fit the model using data for 10 accident years as well as 18 accident years.

Recall that data from the first 8 of the 18 accident years, contain additional claims that are fully developed. We note from Table \ref{T:Dparest} that two sets of data do lead to different estimates. Such difference is much more pronounced for parameter $a$'s than $\phi$'s. If the 8 additional years of data is considered to be representative of the recent 10 years, it would make sense to rely on the estimates based on the 18 years data. On the other hand, if there have been significant operational changes in the company in the recent 10 years compared to the previous 8 years, it would be sensible to rely on the estimates based on recent 10 years of data.

We note that the estimated value of $a_0 (=a_1+\cdots+a_n)$ is $\approx~4512$ based on 10 years of data and is $\approx~1143$ based on 18 years of data. In both cases, the ratio $a_0/(a_0+1)$ is close to 1. In addition, we have checked empirically for several companies, that for estimation based on 10 or 18 accident years, it is indeed the case that $a_0/(a_0+1)\approx 1$. As noted in Section \ref{subsec:mle}, this means that the prediction from the Dirichlet model is expected to be close to the prediction from the Chain-Ladder method.  Below we use the Chain-Ladder method as a benchmark to compare the results. Although the Chain-Ladder predictions are not based on a statistical model, \cite{Mack1993} proposed a distribution-free approach to compute standard errors for the Chain-Ladder predictions based on its implicit assumptions. Specifically, we use Mack's Chain-Ladder approach as our benchmark.

Table \ref{T:LDF} presents the estimated loss development factors from the Dirichlet Model and that obtained from the Mack's Chain-Ladder approach.  The point estimates from the Dirichlet model are close to those from the traditional Chain-Ladder. Recall that the definition of year-to-year development factors $\gamma_{k:k+1}$ in the proposed Dirichlet model is consistent with the traditional definition used in \cite{Mack1993}. However, the underlying model assumptions and estimation strategies are different. Hence it is satisfying to note that our approach leads to similar point estimates, although the standard errors are different. In addition, we emphasize that even when $a_0/(a_0+1)\approx 1$, the estimated development factors from the Dirichlet and Chain-Ladder methods are not identical as shown in Table \ref{T:LDF}. The reported estimates of development factors for the Dirichlet model in Table \ref{T:LDF} are based on the MLEs of model parameters.
\begin{table}[htbp]
  \centering
  \caption{\label{T:Dparest}MLE of parameters in the Dirichlet model }
  \resizebox{\textwidth}{!}{
    \begin{tabular}{|crr|rr|crr|rr|}
          \hline
    Parameter & \multicolumn{2}{p{4cm}|}{Dirichlet Model             } & \multicolumn{2}{p{4cm}|}{Dirichlet Model} &  & \multicolumn{2}{p{4cm}|}{Dirichlet Model} & \multicolumn{2}{p{4cm}|}{Dirichlet Model} \\
      & \multicolumn{2}{p{4cm}|}{(10 accident years)} & \multicolumn{2}{p{4cm}|}{(18 accident years)} & Parameter & \multicolumn{2}{p{4cm}|}{(10 accident years)} & \multicolumn{2}{p{4cm}|}{(18 accident years)} \\
    \hline
          & \multicolumn{1}{c}{Estimate} & \multicolumn{1}{c|}{SE} &  \multicolumn{1}{c}{Estimate} & \multicolumn{1}{c|}{SE}    &       & \multicolumn{1}{c}{Estimate} & \multicolumn{1}{c|}{SE} & \multicolumn{1}{c}{Estimate} & \multicolumn{1}{c|}{SE} \\
    $a_1$    &     1,293.81  &                326.87  &                       347.61  &    48.74  & $\phi_9$ &               0.629  &             0.000  &             0.629  &         0.001  \\
    $a_2$    &     1,006.78  &                254.81  &                       269.54  &    37.64  & $\phi_{10}$ &               0.719  &             0.002  &             0.716  &         0.003  \\
    $a_3$    &        644.73  &                164.21  &                       166.12  &    23.46  & $\phi_{11}$ &               0.766  &             0.004  &             0.759  &         0.005  \\
    $a_4$    &        497.13  &                126.16  &                       126.00  &    17.65  & $\phi_{12}$ &               0.774  &             0.005  &             0.761  &         0.006  \\
    $a_5$    &        338.73  &                  86.17  &                         82.46  &    11.67  & $\phi_{13}$ &               0.773  &             0.006  &             0.753  &         0.008  \\
    $a_6$    &        249.80  &                  63.46  &                         56.37  &       8.19  & $\phi_{14}$ &               0.745  &             0.007  &             0.720  &         0.010  \\
    $a_7$    &        186.01  &                  48.06  &                         38.40  &       5.69  & $\phi_{15}$ &               0.758  &             0.009  &             0.728  &         0.014  \\
    $a_8$    &        138.62  &                  36.27  &                         27.52  &       4.29  & $\phi_{16}$ &               0.725  &             0.011  &             0.691  &         0.017  \\
    $a_9$    &           93.51  &                  25.25  &                         17.63  &       2.84  & $\phi_{17}$ &               0.766  &             0.016  &             0.724  &         0.024  \\
    $a_{10}$   &           63.16  &                  19.51  &                         12.06  &       2.15  & $\phi_{18}$ &               0.682  &             0.022  &             0.643  &         0.034  \\

    \hline
    \end{tabular}}%
    ~\\~\\
\text{\small Note: For brevity, the estimates of $\phi_1,\ldots, \phi_8$ corresponding to years with fully developed claims are not shown. }
\end{table}%

\begin{table}[htbp]
  \centering
  \caption{\label{T:LDF} MLE of loss development factors associated with the Dirichlet model, and based on an industry benchmark}
  \resizebox{0.9\textwidth}{!}{
    \begin{tabular}{|c|cc|cc|cc|cc|}
    \hline
    Development  & \multicolumn{2}{p{4cm}|}{Dirichlet Model               } & \multicolumn{2}{p{4cm}|}{Dirichlet Model} & \multicolumn{2}{p{4cm}|}{Mack Chain-Ladder } & \multicolumn{2}{p{4cm}|}{Mack Chain-Ladder} \\
     Factor  & \multicolumn{2}{p{4cm}|}{(10 accident years)} & \multicolumn{2}{p{4cm}|}{(18 accident years)} & \multicolumn{2}{p{4cm}|}{(10 accident years)} & \multicolumn{2}{p{4cm}|}{(18 accident years)} \\
     \hline
          & Estimate & SE    & Estimate & SE    & \multicolumn{1}{c}{Estimate} & \multicolumn{1}{c|}{SE} & \multicolumn{1}{c}{Estimate} & \multicolumn{1}{c|}{SE} \\
    $\gamma_{1:2}$ &           1.778  &                0.0138  &                         1.775  &  0.0175  &           1.779  &             0.0087  &             1.781  &             0.013  \\
     $\gamma_{2:3}$  &           1.280  &                0.0055  &                         1.269  &  0.0072  &           1.281  &             0.0082  &             1.269  &             0.006  \\
     $\gamma_{3:4}$  &           1.169  &                0.0040  &                         1.161  &  0.0049  &           1.169  &             0.0039  &             1.160  &             0.003  \\
     $\gamma_{4:5}$  &           1.098  &                0.0029  &                         1.091  &  0.0034  &           1.098  &             0.0024  &             1.090  &             0.003  \\
      $\gamma_{5:6}$ &           1.066  &                0.0025  &                         1.057  &  0.0026  &           1.066  &             0.0012  &             1.057  &             0.002  \\
      $\gamma_{6:7}$  &           1.046  &                0.0023  &                         1.037  &  0.0020  &           1.046  &             0.0019  &             1.037  &             0.003  \\
    $\gamma_{7:8}$  &           1.033  &                0.0021  &                         1.025  &  0.0017  &           1.033  &             0.0026  &             1.025  &             0.002  \\
    $\gamma_{8:9}$ &           1.021  &                0.0020  &                         1.016  &  0.0014  &           1.022  &             0.0013  &             1.016  &             0.002  \\
   $\gamma_{9:10}$  &           1.014  &                0.0023  &                         1.011  &  0.0012  &           1.014  &             0.0008  &             1.010  &             0.001  \\
      \hline
    \end{tabular}}%
  \label{tab:addlabel}%
\end{table}%

\subsection{Prediction} \label{subsec:pred}
Using the bootstrap method in \ref{subsec:mle}, we derive the predictive distribution of accident-year loss reserves for the selected insurer.
We summarize in Table \ref{T:ULforecast} the forecasted loss ratios and the 95\% prediction intervals at the 10th development year, based on the model fitted to 10 accident years, as well as model fitted to 18 accident years of data. As a benchmark, we present the corresponding forecasts from the Chain-Ladder method along with the Dirichlet model. The actual realized loss ratios are also reported in the table for assessing the predictive performance. There are a couple of important observations from the table. First, both the point predictions and the 95\% prediction intervals for the Dirichlet model are comparable to that obtained from the Chain-Ladder method. This is as anticipated from the theoretical results in Section \ref{sec:inference} that without additional information on the loss development patterns, reserve predictions from the MLEs of the Dirichlet model lead to Chain-Ladder forecasts under certain conditions, although the prediction intervals are derived from different mechanisms and can be quite different from each other.

Second, for this particular insurer, the predictions with additional 8 years of fully developed claims data appear to be much improved compared with the prediction using the traditional 10-year triangle data. For most accident years, the predicted values from 18 years of data are closer to the actual values than the forecasts from 10 years of data. It is further noted that the 95\% prediction intervals from 18 years of data happen to have
captured the true loss ratios for all accident years. This is not the case for the predictions based on 10 years of data which could lead to serious over-reserving in most recent years. We expressly note that in general it may not be appropriate to use additional years of data if they are not representative of the recent years, e.g. if there have been significant changes in the operations of the company. However, further examination of the selected insurer's operation supports the usage of the additional data. Specifically, the insurer has been writing the worker's compensation business from a regional market for over hundred years with stable underwriting criterion and business mix in the portfolio. More importantly, the insurer focuses on coverage in assigned risk markets that serve as a safety net for employers that are unable to obtain workers compensation insurance from a ``regular'' insurer because of their poor or less credible loss history. If one thinks of assigned risks as ``bad risks'', it is intuitive to understand that it will require more data to capture the inherent higher uncertainty in the prediction.
\begin{table}[htbp]
  \centering
  \caption{\label{T:ULforecast} Actual and forecasted cumulative loss ratios at the end of the 10th development year}
  \resizebox{\textwidth}{!}{
    \begin{tabular}{|p{1.5cm}|p{3cm}|cc|cc|cc|cc|}
    \hline
    Accident  & Actual Loss Ratio & \multicolumn{2}{p{4cm}|}{Dirichlet Model } & \multicolumn{2}{p{4cm}|}{Dirichlet Model } & \multicolumn{2}{p{4cm}|}{Mack Chain-Ladder} & \multicolumn{2}{p{4cm}|}{Mack Chain-Ladder} \\
 Year   & at  10th Dev Year & \multicolumn{2}{p{4cm}|}{(10 accident years data)} & \multicolumn{2}{p{4cm}|}{(18 accident years data)} & \multicolumn{2}{p{4cm}|}{(10 accident years data)} & \multicolumn{2}{p{4cm}|}{ (18 accident years data)} \\
    \hline
          & Actual   & Predicted & 95\% Interval & Predicted & 95\% Interval & Predicted & 95\% Interval & Predicted & 95\% Interval \\
     1997  &                      0.629  &         0.629  &  [0.629,0.629]  &            0.629  &  [0.629,0.629]  &                          0.629  &  [0.629,0.629]  &                  0.629  &  [0.629,0.629]  \\
    1998  &                      0.719  &         0.718  &  [0.714,0.723]  &            0.715  &  [0.709,0.72]  &                          0.719  &  [0.717,0.721]  &                  0.716  &  [0.710,0.722]  \\
    1999  &                      0.763  &         0.765  &  [0.758,0.772]  &            0.758  &  [0.749,0.765]  &                          0.766  &  [0.762,0.770]  &                  0.759  &  [0.749,0.769]  \\
    2000  &                      0.767  &         0.774  &  [0.765,0.783]  &            0.761  &  [0.747,0.772]  &                          0.774  &  [0.766,0.782]  &                  0.761  &  [0.745,0.777]  \\
    2001  &                      0.765  &         0.772  &  [0.760,0.784]  &            0.753  &  [0.736,0.767]  &                          0.773  &  [0.763,0.783]  &                  0.753  &  [0.733,0.773]  \\
    2002  &                      0.741  &         0.745  &  [0.730,0.759]  &            0.719  &  [0.698,0.739]  &                          0.745  &  [0.735,0.755]  &                  0.720  &  [0.696,0.744]  \\
    2003  &                      0.722  &         0.758  &  [0.742,0.776]  &            0.727  &  [0.699,0.752]  &                          0.759  &  [0.745,0.773]  &                  0.727  &  [0.700,0.754]  \\
    2004  &                      0.705  &         0.725  &  [0.704,0.747]  &            0.691  &  [0.657,0.723]  &                          0.726  &  [0.706,0.746]  &                  0.690  &  [0.659,0.721]  \\
    2005  &                      0.729  &         0.766  &  [0.734,0.796]  &            0.723  &  [0.676,0.771]  &                          0.767  &  [0.732,0.802]  &                  0.722  &  [0.681,0.763]  \\
    2006  &                      0.629  &         0.681  &  [0.638,0.723]  &            0.642  &  [0.576,0.708]  &                          0.683  &  [0.644,0.722]  &                  0.644  &  [0.587,0.701]  \\
    \hline
    \end{tabular}}%
\end{table}%

\begin{figure}
\caption{\label{fig:D18} Actual and forecasted cumulative loss ratios by accident year.}
   \includegraphics[width=150mm]{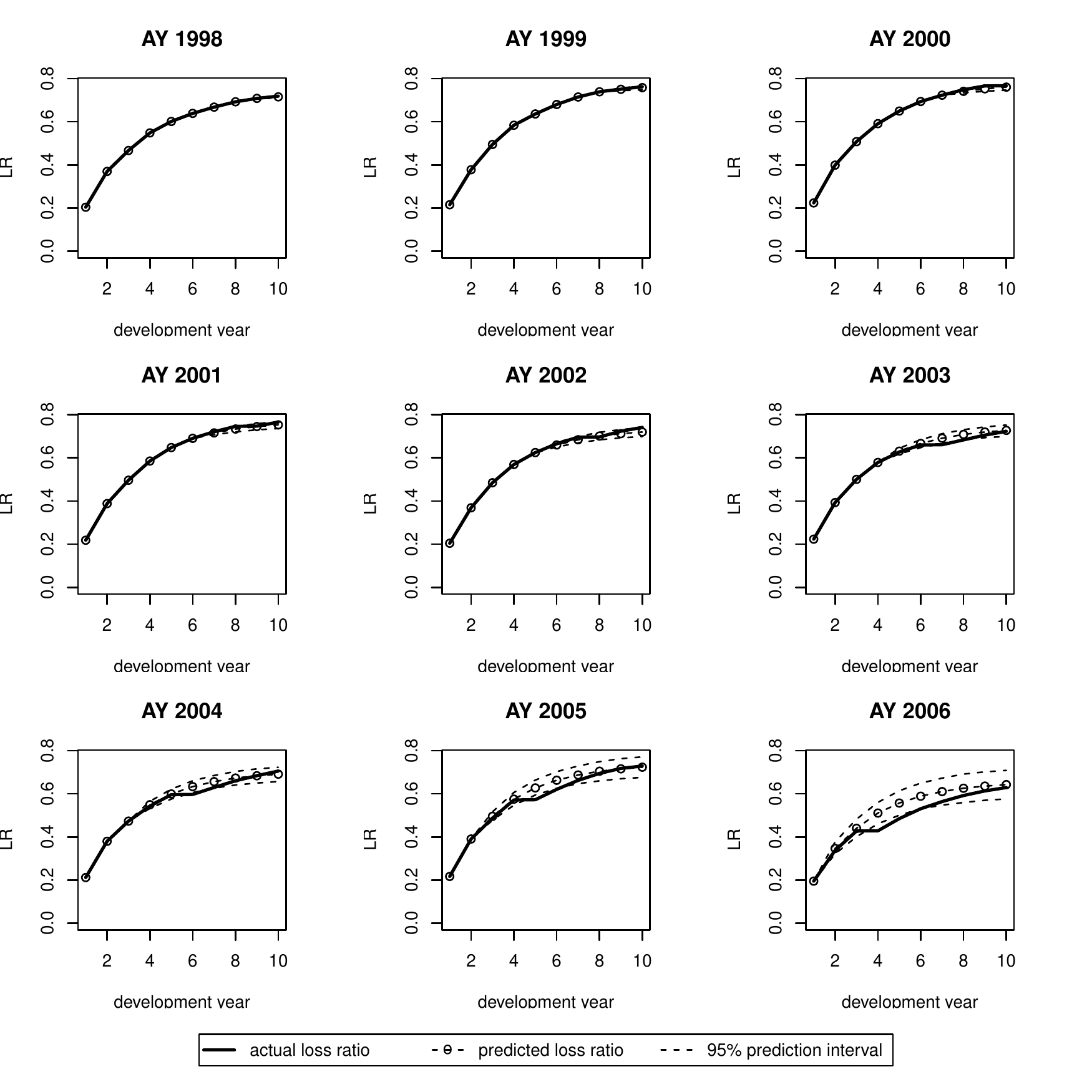}
\end{figure}

Finally, it is worth stressing one advantage of the Dirichlet model over the Mack's Chain-Ladder approach. To quantify the predictive uncertainty, the Chain-Ladder approach relies on the conditional mean squared error of prediction and constructs the prediction interval using normal approximation. In contrast, the proposed model-based approach allows us to derive not only an interval estimate, but also the entire predictive distribution for the unpaid losses. Furthermore, the predictive distribution can be obtained for any outcome of interest, be it incremental paid losses, accident year reserves, or calendar year reserves, etc. The Chain-Ladder approach does not enjoy this flexibility in terms of calculating the prediction error. For illustration, we show in Figure \ref{fig:D18} the forecasted paths (both point and interval predictions) of cumulative loss ratios by accident year along with the actual loss ratios. A larger prediction interval is observed when one has fewer historical data or one predicts further into the future.

\subsection{Validation} \label{subsec:val}
The findings from the previous section are interesting but based on one particular company. To check whether similar conclusions hold in general, we carry out the analysis for the 139 large insurers in the NAIC database selected based on criteria described in Section \ref{subsec:sampling}.

We start by applying the goodness-of-fit testing procedure described in Section \ref{subsec:GOF} to each of the 139 insurers. We carry out the test at 5\% level, separately  considering data from 10 recent accident years as well as 18 accident years. When we carry out the test based on data from 10 accident years, the Dirichlet model is supported for 112 companies out of 139.  Similarly, when we carry out the test using data for 18 accident years, the Dirichlet model is supported for 76 companies out of 139. This suggests that the Dirichlet model seems to be a reasonable fit for a large number of companies. Further, the model usually fits better when we consider only the recent 10 accident years, as against trying to fit it to data from 18 accident years. So, a common model may not always be appropriate for the extended data consisting 18 accident years. The testing procedure could be used as one guideline to determine whether or not to use extended 18 years of data while developing forecasts based on this model for a given company.

Next, we obtain predictions for the selected 139 companies using the Dirichlet model as well as the Mack's Chain-Ladder approach. We reiterate that the predictions are assessed using the the actual loss ratios in the hold-out sample. We compute three metrics by each accident year using both 10-year and 18-year data: (i) Root mean squared error (RMSE), which is defined as the square root of average (over 139 companies) squared deviations of actual loss ratios from predicted loss ratios; (ii) Coverage of the 95 \% prediction interval (Cov95), which is the percentage of companies out of the 139 insurers for which the 95\% prediction interval contained the true loss ratio; and (iii) Average length of the 95\% prediction interval (Len95), which is the average length (across 139 companies) of the prediction interval. The results are summarized in Table \ref{T:overallacc}. Ideally, the coverage should be approximately 0.95, i.e. the 95\% prediction interval captures the true value for 95\% of the time. For both the RMSE and the length of the 95\% prediction interval, when other things equal, the lower the value, the better is the prediction.


First, we compare results from the Dirichlet model and the Mack's Chain-Ladder approach. The RMSE for the two methods are comparable for both 10-year data and 18-year data cases. This further supports our initial observation following \textit{Proposition 1} that the predictions from the Dirichlet model, when there is no additional information, will be close to the Chain-Ladder predictions. The metrics based on prediction intervals suggest that the Dirichlet model outperforms the Mack's Chain-Ladder approach. For the case of 10-year data, the coverage from the Chain-Ladder prediction is further below the target value than that from the Dirichlet prediction, especially for early accident years. The lower coverage also explains the smaller length of the prediction interval from the Chain-Ladder method. For the case of 18-year data, the coverage of the Chain-Ladder prediction is better but at the cost of inflating the length of the prediction interval. As a result, the Chain-Ladder method shows wider prediction interval yet still smaller coverage compared to the Dirichlet model. Furthermore, the comparison also suggests that the Dirichlet prediction is more consistent in its accuracy across accident years than the Chain-Ladder prediction.

Second, we compare predictions using 10 years of data and 18 years of data. For both Dirichlet and Chain-Ladder methods, the usage of additional 8 years of claims data inflates the RMSE for most accident years. In addition, when using additional data, the coverage of the 95\% prediction interval from the Dirichlet model becomes worse, while the average length of the prediction interval remains at similar level. For the Chain-Ladder method, the additional 8 years of data help improve the coverage of the prediction interval, but the bigger price paid is the resulting larger length of the prediction interval. Overall, the results in Table \ref{T:overallacc} indicate that this particular dataset does not support the use of 18-year data in general. One explanation is that many of the selected insurers might experience significant changes in the operations so that the additional 8 years of data are not representative of the recent 10 years of observations in learning the claim development patterns over time.

\begin{table}[htbp]
  \centering
  \caption{\label{T:overallacc} Prediction accuracy measures for the selected 139 companies}
  \resizebox{\textwidth}{!}{
    \begin{tabular}{|c|ccc|ccc|ccc|ccc|}
    \hline
    Accident& \multicolumn{3}{p{4.5cm}|}{Dirichlet Model                        } & \multicolumn{3}{p{4.5cm}|}{Dirichlet Model} & \multicolumn{3}{p{4.5cm}|}{Mack Chain-Ladder} & \multicolumn{3}{p{4.5cm}|}{Mack Chain-Ladder                    } \\
       Year & \multicolumn{3}{p{4.5cm}|}{ (10 accident years data)} & \multicolumn{3}{p{4.5cm}|}{ (18 accident years data)} & \multicolumn{3}{p{4.5cm}|}{(10 accident years data)} & \multicolumn{3}{p{4.5cm}|}{(18 accident years data)} \\
        \hline
          & \multicolumn{1}{p{1.5cm}}{RMSE} & \multicolumn{1}{p{1.5cm}}{Cov95} & \multicolumn{1}{p{1.5cm}|}{Len95} & \multicolumn{1}{p{1.5cm}}{RMSE} & \multicolumn{1}{p{1.5cm}}{Cov95} & \multicolumn{1}{p{1.5cm}|}{Len95} & \multicolumn{1}{p{1.5cm}}{RMSE} & \multicolumn{1}{p{1.5cm}}{Cov95} & \multicolumn{1}{p{1.5cm}|}{Len95} & \multicolumn{1}{p{1.5cm}}{RMSE} & \multicolumn{1}{p{1.5cm}}{Cov95} & \multicolumn{1}{p{1.5cm}|}{Len95} \\
    1997  &                -    &           1.000  &                 -    &                -    &         1.000  &                    -    &                -    &         1.000  &                -    &                -    &         1.000  &                -    \\
    1998  &         0.007  &           0.950  &          0.039  &         0.006  &         0.928  &             0.034  &         0.007  &         0.446  &         0.011  &         0.005  &         0.806  &         0.018  \\
    1999  &         0.052  &           0.871  &          0.082  &         0.041  &         0.813  &             0.064  &         0.062  &         0.496  &         0.029  &         0.042  &         0.640  &         0.035  \\
    2000  &         0.059  &           0.871  &          0.099  &         0.062  &         0.748  &             0.084  &         0.068  &         0.619  &         0.045  &         0.062  &         0.612  &         0.064  \\
    2001  &         0.053  &           0.813  &          0.109  &         0.059  &         0.705  &             0.098  &         0.056  &         0.712  &         0.061  &         0.061  &         0.619  &         0.159  \\
    2002  &         0.057  &           0.763  &          0.114  &         0.073  &         0.705  &             0.108  &         0.056  &         0.691  &         0.074  &         0.069  &         0.576  &         0.221  \\
    2003  &         0.050  &           0.791  &          0.119  &         0.069  &         0.727  &             0.120  &         0.042  &         0.770  &         0.092  &         0.059  &         0.691  &         0.275  \\
    2004  &         0.059  &           0.791  &          0.137  &         0.089  &         0.719  &             0.150  &         0.061  &         0.856  &         0.134  &         0.089  &         0.741  &         0.335  \\
    2005  &         0.120  &           0.806  &          0.207  &         0.171  &         0.741  &             0.219  &         0.116  &         0.871  &         0.186  &         0.132  &         0.791  &         0.409  \\
    2006  &         0.219  &           0.871  &          0.437  &         0.340  &         0.871  &             0.479  &         0.219  &         0.835  &         0.582  &         0.236  &         0.856  &         0.737  \\
    \hline
    \end{tabular}}%
\end{table}%

\subsection{Bayesian Inference}\label{sec:bayesinference}
Here, we carry out Bayesian inference for the same selected insurer analyzed in subsections \ref{subsec:est} and \ref{subsec:pred}. An important feature of the proposed Dirichlet model is that it gives a formal statistical approach such that one could view both the Chain-Ladder and Bornhuetter-Ferguson methods in a unified modeling framework. As demonstrated in the previous subsections,  the Dirichlet model leads to predictions similar to the Chain-Ladder method, when there is no additional information beyond the 10 year (or 18 year) triangular data of claims. The Bornhuetter-Ferguson approach necessarily requires additional (internal or external) information on the ultimate loss ratio to obtain the reserve prediction.
Since our approach is a statistical model that  nests the Bornhuetter-Ferguson approach, such additional information can be incorporated through a Bayesian framework by assuming a suitable prior on the Dirichlet model parameters.

Recall that in previous sections we interpret parameter $\phi_i$ in the Dirichlet model as the expected ultimate loss ratio for accident year $i$. A natural extension using a Bayesian framework is to consider a hierarchical prior specification for $\phi_i$ to allow for borrowing of information across accident years. Specifically, we take
\[\phi_1, \phi_2, \ldots, \phi_m \stackrel{iid}{\sim} uniform(0, \phi), \mbox{ with  } p(\phi)\propto 1. \]
The hierarchical prior on $\phi_i$ assumes that the changes in the risks undertaken by the company across years are purely random fluctuations and not due to a systematic shift in risk profile. We further assume a flat prior for all the other unknown parameters, i.e,
\[p(a_1, a_2, \ldots, a_n, b_n) \propto 1. \]

Under the hierarchical framework, we consider the following three models:
 \begin{itemize}
\item[] (i) Based only on the recent 10 accident years of data;
\item[] (ii) Based on 18 accident years of data, of which the claims in the additional 8 accident years are fully developed; 
\item[] (iii) Based on 18 accident years of data, but in addition, we impose a constraint on the expected tail loss ratio beyond $n$ years ($n=10$), i.e. $$E\left[1-\frac{S_{i,1:n}}{\phi_i} \right]=\frac{b_n}{a_0 +b_n}\geq \alpha.$$ 
\end{itemize}
The above three models illustrate different levels of prior information that an analyst could infuse into the Bayesian inference. Model (i) represents the basic hierarchical specification which notably differs from the Dirichlet model in Sections \ref{subsec:est}-\ref{subsec:val} that does not allow learning across accident years. Model (ii) is based on the assumption that the additional fully developed claims in early accident years are representative of most recent 10 accident years, and hence contribute to the learning of the model parameters in the Bayesian specification. Because worker's compensation is a long-tailed line of business, claims are expected to further develop after 10 years. Model (iii) allows us to incorporate prior knowledge on tail development into inference.

We formulate the Bayesian estimation as in Section  \ref{subsec:bayes} and implement it using R-Stan \citep{rstan2018}. For the same reasons as mentioned in Section \ref{subsec:mle}, we impose the constraint $b_n\geq 1$. We note that the Markov Chain Monte Carlo simulations tend to diverge without this condition. In case (iii), a flat prior on $b_n$ indicates a support on interval $[ \frac{\alpha}{1-\alpha}a_0,\infty)$. Industry benchmark studies (see, for instance, \cite{ShermanDiss2005}) suggest that $\alpha\approx 19\%$, which translates to the constraint $b_n\geq 0.24 a_0$.

\begin{figure}
\caption{\label{fig:predmethods} Comparison of predicted loss ratios at the 10th development year from different methods}
   \includegraphics[height=10cm, width=16cm]{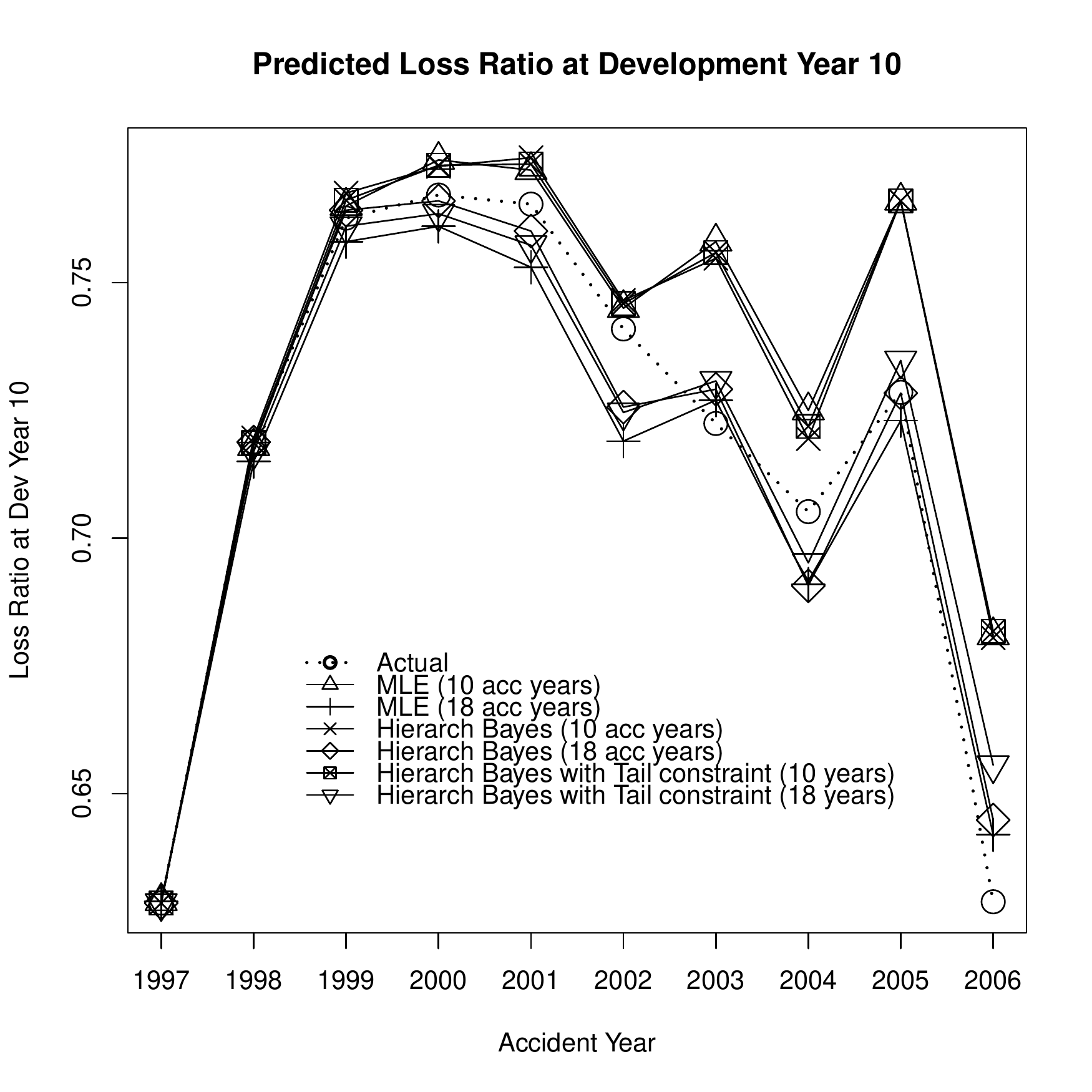}
\end{figure}

We apply the Bayesian formulation to the same insurer analyzed in Sections \ref{subsec:est} and \ref{subsec:pred}.
Figure \ref{fig:predmethods} compares predicted loss ratio at the end of the 10th development year by accident year, using different estimation methods. The figure suggests that the predictions from the MLE and hierarchical Bayesian specifications using 18 years of data are closer to the actual loss ratio, especially for the most recent three accident years, than the predictions using 10 years of data. This observation is consistent with the conclusions that we drew in Section \ref{subsec:pred}.

For a closer comparison between various models, we summarize in Table \ref{T:BayesPred} three metrics for predictions from the MLE and Bayesian formulations based on 10 as well as 18 accident years of data. The actual loss ratio is also given for reference. For each accident year, we calculate (i) the absolute deviation of the predicted loss ratio from the actual value; (ii) whether the 95\% prediction interval captures the actual loss ratio (1/0); and (iii) the length of the 95\% prediction interval. The overall average (across accident years) of the three metrics is reported at the bottom of the table. First, both average deviation and average coverage suggest predictions based 18 years of data are better than those based on 10 years of data, confirming the result observed in Figure \ref{fig:predmethods}. Second, the prediction intervals from the MLE are in general wider than that from the Bayesian formulations which is as anticipated because of the learning effect across accident years. Third and most important, the hierarchical formulation based on 18 years of data with tail constraint turns out to provide the best prediction, i.e. the prediction intervals capture the true loss ratio for all accident years, and in the meanwhile the intervals are narrower compared to other methods. We note that the information on tail factor cannot be learnt and validated from the available data (i.e. Table \ref{T:premiumlossdata}) and needs to be necessarily provided as an additional input. Hence, it is an important feature of the model that such external information can be systematically incorporated.

\begin{table}[htbp]
 \caption{\label{T:BayesPred} Comparison of predictions from the MLE and Bayesian methods.}
\begin{subtable}
  \centering
  \resizebox{\textwidth}{!}{
    \begin{tabular}{|p{1cm}p{1cm}|p{2cm}p{1.5cm}p{1.5cm}p{1.5cm}|p{2cm}p{1.5cm}p{1.5cm}p{1.5cm}|p{2cm}p{1.5cm}p{1.5cm}p{1.5cm}|}
    \hline
          &       & \multicolumn{4}{p{6.5cm}|}{MLE Dirichlet Model                                                                     } & \multicolumn{4}{p{6.5cm}|}{Hierarchical Bayes                                                                                } & \multicolumn{4}{p{6.5cm}|}{Hierarch Bayes with Tail Constraint                                            } \\

                    &       & \multicolumn{4}{p{6.5cm}|}{ (10 accident years data)} & \multicolumn{4}{p{6.5cm}|}{ (10 accident years data)} & \multicolumn{4}{p{6.5cm}|}{ (10 accident years data)} \\
                      \hline
    Acc Year & Actual Loss Ratio & 95\% Interval & $|$Actual-Predicted$|$& Interval contains actual & Interval Length & 95\% Interval & $|$Actual-Predicted$|$& Interval contains actual & Interval Length & 95\% Interval & $|$Actual-Predicted$|$& Interval contains actual & Interval Length \\
  \hline
      1997  &         0.629  &  [0.629,0.629]  &                             -    &                           1  &                  -    &  [0.629,0.629]  &                      0.000  &                       1  &                     -    & [0.629,0.629] &                    0.000  & 1     &                             -     \\
    1998  &         0.719  &  [0.714,0.723]  &                      0.001  &                           1  &           0.009  &  [0.719,0.72]  &                      0.001  &                       1  &              0.001  & [0.716,0.721] &                    0.000  & 1     &                       0.005  \\
    1999  &         0.763  &  [0.758,0.772]  &                      0.002  &                           1  &           0.014  &  [0.767,0.768]  &                      0.005  &                     -    &              0.001  & [0.762,0.77] &                    0.003  & 1     &                       0.008  \\
    2000  &         0.767  &  [0.765,0.783]  &                      0.007  &                           1  &           0.018  &  [0.771,0.774]  &                      0.006  &                     -    &              0.003  & [0.768,0.779] &                    0.006  & 0     &                       0.011  \\
    2001  &         0.765  &  [0.76,0.784]  &                      0.007  &                           1  &           0.024  &  [0.772,0.777]  &                      0.009  &                     -    &              0.005  & [0.766,0.78] &                    0.008  & 0     &                       0.014  \\
    2002  &         0.741  &  [0.73,0.759]  &                      0.004  &                           1  &           0.029  &  [0.743,0.75]  &                      0.006  &                     -    &              0.007  & [0.738,0.755] &                    0.005  & 1     &                       0.017  \\
    2003  &         0.722  &  [0.742,0.776]  &                      0.036  &                          -    &           0.034  &  [0.75,0.76]  &                      0.033  &                     -    &              0.010  & [0.745,0.767] &                    0.034  & 0     &                       0.022  \\
    2004  &         0.705  &  [0.704,0.747]  &                      0.020  &                           1  &           0.043  &  [0.713,0.727]  &                      0.014  &                     -    &              0.014  & [0.709,0.736] &                    0.017  & 0     &                       0.027  \\
    2005  &         0.729  &  [0.734,0.796]  &                      0.037  &                          -    &           0.062  &  [0.755,0.777]  &                      0.037  &                     -    &              0.022  & [0.746,0.786] &                    0.037  & 0     &                       0.040  \\
    2006  &         0.629  &  [0.638,0.723]  &                      0.052  &                          -    &           0.085  &  [0.663,0.698]  &                      0.052  &                     -    &              0.035  & [0.654,0.712] &                    0.053  & 0     &                       0.058  \\
    \hline
    Average &       &       &                      0.017  &                   0.700  &           0.032  &       &                      0.016  &              0.200  &              0.010  &       &                    0.016  &              0.400  &                       0.020  \\
    \hline
    \end{tabular}}%
    \end{subtable}
    ~\\
    ~\\
\begin{subtable}
  \centering
  \resizebox{\textwidth}{!}{
    \begin{tabular}{|p{1cm}p{1cm}|p{2cm}p{1.5cm}p{1.5cm}p{1.5cm}|p{2cm}p{1.5cm}p{1.5cm}p{1.5cm}|p{2cm}p{1.5cm}p{1.5cm}p{1.5cm}|}
    \hline
          &       & \multicolumn{4}{p{6.5cm}|}{MLE Dirichlet Model                                                                     } & \multicolumn{4}{p{6.5cm}|}{Hierarchical Bayes                                                                                } & \multicolumn{4}{p{6.5cm}|}{Hierarch Bayes with Tail Constraint                                            } \\

                    &       & \multicolumn{4}{p{6.5cm}|}{ (18 accident years data)} & \multicolumn{4}{p{6.5cm}|}{ (18 accident years data)} & \multicolumn{4}{p{6.5cm}|}{ (18 accident years data)} \\
                      \hline
    Acc Year & Actual Loss Ratio & 95\% Interval & $|$Actual-Predicted$|$& Interval contains actual & Interval Length & 95\% Interval & $|$Actual-Predicted$|$ & Interval contains actual & Interval Length & 95\% Interval &$|$Actual-Predicted$|$ & Interval contains actual & Interval Length \\
  \hline
     1997  &         0.629  &  [0.629,0.629]  &                             -    &                           1  &                  -    &  [0.629,0.629]  &                      0.000  &                       1  &                     -    & [0.629,0.629] &                    0.000  & 1     &                             -     \\
    1998  &         0.719  &  [0.709,0.72]  &                      0.004  &                           1  &           0.011  &  [0.718,0.719]  &                      0.000  &                       1  &              0.001  & [0.713,0.722] &                    0.002  & 1     &                       0.009  \\
    1999  &         0.763  &  [0.749,0.765]  &                      0.005  &                           1  &           0.016  &  [0.763,0.765]  &                      0.001  &                       1  &              0.002  & [0.754,0.769] &                    0.002  & 1     &                       0.015  \\
    2000  &         0.767  &  [0.747,0.772]  &                      0.006  &                           1  &           0.025  &  [0.764,0.768]  &                      0.001  &                       1  &              0.004  & [0.754,0.774] &                    0.004  & 1     &                       0.020  \\
    2001  &         0.765  &  [0.736,0.767]  &                      0.012  &                           1  &           0.031  &  [0.757,0.763]  &                      0.005  &                     -    &              0.006  & [0.745,0.771] &                    0.008  & 1     &                       0.026  \\
    2002  &         0.741  &  [0.698,0.739]  &                      0.022  &                          -    &           0.041  &  [0.722,0.73]  &                      0.015  &                     -    &              0.008  & [0.71,0.74] &                    0.016  & 0     &                       0.030  \\
    2003  &         0.722  &  [0.699,0.752]  &                      0.005  &                           1  &           0.053  &  [0.723,0.735]  &                      0.007  &                     -    &              0.012  & [0.712,0.751] &                    0.009  & 1     &                       0.039  \\
    2004  &         0.705  &  [0.657,0.723]  &                      0.014  &                           1  &           0.066  &  [0.682,0.699]  &                      0.014  &                     -    &              0.017  & [0.671,0.721] &                    0.010  & 1     &                       0.050  \\
    2005  &         0.729  &  [0.676,0.771]  &                      0.006  &                           1  &           0.095  &  [0.715,0.743]  &                      0.001  &                       1  &              0.028  & [0.698,0.776] &                    0.006  & 1     &                       0.078  \\
    2006  &         0.629  &  [0.576,0.708]  &                      0.013  &                           1  &           0.132  &  [0.624,0.667]  &                      0.016  &                       1  &              0.043  & [0.603,0.718] &                    0.027  & 1     &                       0.115  \\
    \hline
    Average &       &       &                      0.009  &                   0.900  &           0.047  &       &                      0.006  &              0.600  &              0.012  &       &                    0.008  &              0.900  &                       0.038  \\
    \hline
    \end{tabular}}%
\end{subtable}
\end{table}%

\section{Concluding Remarks}

In this paper, we propose a novel stochastic loss reserving model for predicting the outstanding liability and quantifying the reserving uncertainty for property-casualty insurers. The mathematical characterization of this model makes it a natural choice for the loss reserving context.  Our main contribution is not just the stochastic model itself but also the new perspective that allows us to view the two industry benchmarks, the Chain-Ladder method and the Bornhuetter-Ferguson method, in a unified modeling framework. We have shown that the Dirichlet model could lead to either Chain-Ladder or Bornhuetter-Ferguson prediction, depending on the available information used in model inference. Specifically, the prediction based on MLE nests the Chain-Ladder prediction and the prediction based on Bayesian estimation with informative priors nests the Bornhuetter-Ferguson prediction. It is well known that the prediction for reserves from the two industry benchmarks are connected but are supported by separate stochastic models. The new perspective provided by the Dirichlet model transforms the selection between two methods to an inference issue.

We stress that the prediction from the proposed Dirichlet model goes far beyond the Chain-Ladder and Bornhuetter-Ferguson methods. First, the accident-year reserves turn out to be a credibility weighted average of predictions from the Chain-Ladder and the expected methods, and the credibility weight is a function of the riskiness of claims. Second, the prediction for ultimate losses featured an embedded tail development factor which allows the analyst to incorporate prior knowledge of the tail development of claims into model inference.

Another unique feature of our study is to supplement the traditional triangular loss data with additional years of fully developed claims. Using a case study for a particular insurer, we demonstrated how such additional data could contribute to the learning of loss development patterns. In the meanwhile, we also caution that serious bias could be introduced into the prediction if there are significant changes in an insurer's operation such that the additional data are not representative of the most recent data. With the advantage of the proposed Bayesian inference, one potential for future research is to develop some informative priors on the loss development factors using the additional fully developed claims, and then use the informative priors in the hierarchical specification of the Dirichlet model.

\bibliographystyle{chicago}
\bibliography{ref_TriangleReserve1}

\newpage

\section*{Appendices}

\subsection*{A.1 Dirichlet Distribution}

Let $\bm{P}=(P_1,\ldots,P_K)$ be a random vector with $K\ge 2$ components. Then $\bm{P}$ is said to follow the Dirichlet distribution of order $K\ge 2$, which we denote by $\bm{P}=(P_1,\ldots,P_K)\sim {\rm Dir}(\alpha_1,\ldots,\alpha_K)$, if its density is given by:
\begin{align}
f(\bm{p};\alpha_1,\ldots,\alpha_K) = \frac{\Gamma\left(\sum_{k=1}^{K}\alpha_k\right)}{\prod_{k=1}^{K}\Gamma(\alpha_k)} \prod_{k=1}^{K}p_k^{\alpha_k-1},
\end{align}
where $\alpha_1,\ldots,\alpha_K$ are parameters of the distribution with $\alpha_k>0$ for each $k$, and $\bm{p}=(p_1,\ldots,p_K)$ is on the $(K-1)$-dimensional probability simplex, i.e. $\sum_{k=1}^{K}p_k=1$ and $p_k\ge 0$ for $k=1,\ldots,K$. Thus, the Dirichlet distribution can be thought of as a distribution over probability mass functions of length $K$.

We briefly summarize below some useful properties of the Dirichlet distribution that are relevant to loss reserve prediction.

A.1. The mean and variance of $\bm{P}$ are shown as:
\begin{align}
{\rm E}(P_k) = \frac{\alpha_k}{\alpha_0},\quad {\rm Cov}(P_k,P_{k'})=\left\{
                                                                                     \begin{array}{ll}
                                                                                       \cfrac{\alpha_k(\alpha_0-\alpha_k)}{\alpha_0^2(\alpha_0+1)} & {\rm if}~k=k' \\
                                                                                       \cfrac{-\alpha_k\alpha_{k'}}{\alpha_0^2(\alpha_0+1)} & {\rm if}~k\neq k' \\
                                                                                     \end{array},
                                                                                   \right.
\end{align}
where $\alpha_0=\sum_{k=1}^{K}\alpha_k$.

A.2. If $\{B_1,\ldots,B_{l}\}$ is a partition of $\{1,\ldots,K\}$, then
\begin{align}
\left(\sum_{k\in B_1}P_k,\ldots,\sum_{k\in B_l}P_k\right)\sim {\rm Dir}\left(\sum_{k\in B_1}\alpha_k,\ldots,\sum_{k\in B_l}\alpha_k\right).
\end{align}
As a special case, the marginal distribution of $P_k$ is ${\rm Beta}(\alpha_k,\alpha_0-\alpha_k)$ for $k=1,\ldots,K$.

A.3. Let $\bm{P}_{-k}=(P_1,\ldots,P_{k-1},P_{k+1},\ldots,P_K)$ and $\bm{\alpha}_{-k}=(\alpha_1,\ldots,\alpha_{k-1},\alpha_{k+1},\ldots,\alpha_K)$. One can show:
\begin{align}
\frac{1}{1-P_k} (\bm{P}_{-k}|P_k) \sim {\rm Dir}(\bm{\alpha}_{-k}).
\end{align}
\newpage
\subsection*{A.2 Run-off Triangle}
\begin{table}[h]
  \centering
  \caption{\label{T:premiumlossdata} Premium and loss data for a selected insurer (in '000 US dollars)}
  \resizebox{\textwidth}{!}{
    \begin{tabular}{ccccccccccccc}
          &       &       & \multicolumn{10}{c}{Incremental Losses by Development Year} \\
    Year  & Accident Year & Earned Premium & 1     & 2     & 3     & 4     & 5     & 6     & 7     & 8     & 9     & 10 \\
    \hline
    1     & 1989  &                  1,65,339  &             41,891  &             32,156  &             20,520  &             15,256  &               8,170  &             5,317  &             3,415  &             2,504  &             1,967  &      940  \\
    2     & 1990  &                  1,68,293  &             44,050  &             37,311  &             22,339  &             14,356  &               8,419  &             6,258  &             3,545  &             2,981  &             1,468  &  1,265  \\
    3     & 1991  &                  1,83,529  &             47,778  &             39,354  &             21,232  &             16,132  &             10,632  &             6,754  &             4,311  &             2,407  &             1,620  &      993  \\
    4     & 1992  &                  1,92,991  &             49,191  &             42,325  &             22,731  &             16,959  &             11,056  &             6,972  &             4,317  &             2,431  &             2,016  &  1,106  \\
    5     & 1993  &                  2,22,666  &             47,035  &             38,662  &             20,081  &             15,923  &             10,621  &             6,266  &             3,552  &             2,744  &             1,513  &  1,308  \\
    6     & 1994  &                  2,40,844  &             51,538  &             33,518  &             19,964  &             16,713  &             11,076  &             7,526  &             4,835  &             4,450  &             2,273  &  2,155  \\
    7     & 1995  &                  2,58,703  &             46,934  &             31,827  &             21,236  &             15,846  &             11,288  &             6,317  &             5,615  &             4,261  &             2,798  &  2,150  \\
    8     & 1996  &                  2,37,131  &             43,432  &             32,768  &             21,697  &             16,150  &             10,230  &             8,056  &             6,250  &             4,455  &             3,417  &  2,421  \\
\hline
    9     & 1997  &                  2,08,179  &             38,915  &             28,463  &             19,494  &             13,361  &             10,211  &             7,176  &             5,401  &             3,453  &             2,551  &  1,844  \\
    10    & 1998  &                  1,69,361  &             34,596  &             28,089  &             16,409  &             13,813  &               8,966  &             6,333  &             4,913  &             4,196  &             2,670  &  \\
    11    & 1999  &                  1,50,912  &             32,580  &             24,468  &             17,672  &             13,418  &               7,881  &             6,616  &             5,246  &             3,655  &       &  \\
    12    & 2000  &                  1,75,101  &             39,248  &             30,647  &             19,059  &             14,599  &             10,220  &             7,725  &             5,126  &       &       &  \\
    13    & 2001  &                  1,94,483  &             42,433  &             32,981  &             21,082  &             17,274  &             12,151  &             8,309  &       &       &       &  \\
    14    & 2002  &                  2,22,002  &             45,309  &             36,483  &             25,777  &             18,746  &             12,266  &       &       &       &       &  \\
    15    & 2003  &                  2,44,749  &             54,589  &             41,491  &             26,295  &             19,207  &       &       &       &       &       &  \\
    16    & 2004  &                  2,79,994  &             59,399  &             47,007  &             26,169  &       &       &       &       &       &       &  \\
    17    & 2005  &                  3,13,808  &             68,185  &             54,385  &       &       &       &       &       &       &       &  \\
    18    & 2006  &                  3,41,973  &             66,827  &       &       &       &       &       &       &       &       &  \\
    \hline
    \end{tabular}}%
  \label{tab:addlabel}%
\end{table}%

\FloatBarrier
\subsection*{A.3 MLE of the Dirichlet Model}

Using $\cfrac{\partial}{\phi_i}\ln l_i = 0$, it can be seen that for any given values of $\bm{a}$ and $b_n$, the MLE of $\phi_i$ is obtained at:
\begin{align} \label{equ:mlephi}
\hat{\phi}_i = \left\{
           \begin{array}{cc}
             \cfrac{a_0+b_n-1}{a_0} s_{i,1:n}, & 1\leq i \leq m-n \\
             \cfrac{a_0+b_n-1}{\sum_{j=1}^{m+1-i}a_j}s_{i,1:m+1-i}, & m-n+1\leq i \leq m \\
           \end{array},
         \right.
\end{align}
Replacing $\phi_i$ with the above expression in equation (\ref{equ:li}), we subsequently obtain the log-likelihood as a function of $(a_1, a_2,\dots, a_n, b_n)$. Then, taking derivative of the obtained log-likelihood $ll(\bm{a}, b_n)$ with respect to $b_n$, one gets:
\begin{align} \label{equ:mleb}
\frac{\partial}{\partial b_n}ll(\bm{a}, b_n) &= \sum_{i=1}^{m-n} \left\{\Psi(a_0+b_n) -  \Psi(a_0+b_n) + \ln\frac{b_n-1}{a_0+b_n-1}\right\} \nonumber \\
& \quad + \sum_{i=m-n+1}^{m} \left\{\Psi(a_0+b_n) -  \Psi(a_0+b_n-\sum_{j=1}^{m+1-i}a_j) + \ln\frac{a_0+b_n-\sum_{j=1}^{m+1-i}a_j-1}{a_0+b_n-1}\right\},
\end{align}
where $\Psi(\nu)$ is called the digamma function, which is the derivative of log of the gamma function and is given by:
\[\Psi(\nu)= \frac{d}{d\nu} \log\Gamma(\nu), ~~\Gamma(\nu)= \int_{0}^{\infty}x^{\nu-1}e^{-x}dx. \]
Note that equation (\ref{equ:mleb}) can be written as the sum of terms of which each is of the form
\begin{align*}
g_{c}(x)=[\Psi(x+c)-\Psi(x)] - [\ln(x-1+c) - \ln(x-1)], \quad c>0
\end{align*}
It can be verified by computing $g_c(x)$ for a large number of possible values of $c$ and $x$,  that $g_{c}(x)<0$. Therefore, equation(\ref{equ:mleb}) will be negative for any values of $\bm{a}$ and $b_n$. Hence, for any fixed values of $\bm{a}$, the likelihood is a decreasing function of $b_n$ for $b_n\ge 1$. Thus the MLE of $b_n$ is obtained at $\hat{b}_n=1$.

Similarly, if in the log-likelihood expression we replace $\phi_i$ from equation (\ref{equ:mlephi}) and also take $b_n=1$, we express the log-likelihood as a function of $\bm{a}=(a_1, a_2, \ldots, a_n)$. For $l\geq 2$. Denoting $A_l= a_{n-l+2}+ \cdots+a_n$ and recalling $a_0=a_1+ \cdots+a_n$, we show the derivative of the log-likelihood function with respect to $a_j$ as :
\begin{align*}
\frac{\partial}{\partial a_j}ll(\bm{a}) = m\Psi(a_0+1)&-(m-j+1)\Psi(a_j) + \sum_{i=1}^{m-n+1}\log\frac{Y_{in}}{S_{i, 1:n}} + \sum_{i=m-n+2}^{m-j+1} \log \frac{Y_{i, m-i+1}}{S_{i, 1:m-i+1}}\\
& - \sum_{l=n-j+2}^n \Psi(A_l +1) + \sum_{l=2}^{n-j+1}\log\frac{a_0-A_{l}}{a_0} + \sum_{l=n-j+2}^n \log\frac{A_l}{a_0}.
\end{align*}
Let $\Lambda(\nu)=\frac{d}{d\nu} \Psi(\nu)$, i.e. the trigamma function. We can also obtain the second derivative as:
\begin{align*}
\frac{\partial^2}{\partial a_j a_r}ll(\bm{a})= &m\Lambda(a_0+1) -\sum_{l=2}^n \left(I_{(l\geq n-j+2)}\cdot I_{(l\geq n-r+2)} \cdot\Lambda(A_l+1)\right) + (m-j+1) \Gamma(a_j) \cdot I_{(j=r)}\\
&  + \sum_{l=2}^n I_{(l\leq n-j+1)}\cdot \left(\frac{I_{(l\leq n-r+1)}}{a_0-A_l}-\frac{1}{a_0} \right)+ \sum_{l=2}^n I_{(l\geq n-j+2)}\cdot \left( \frac{I_{(l\geq n-r+2)}}{A_l}-\frac{1}{a_0}\right),
\end{align*}
where $I_{(\cdot)}$ denotes the indicator function for the condition in the parenthesis. The gradient vector $\bold{g}(\bm{a})$ and the Hessian matrix $H(\bm{a})$ can be further obtained as:
\[\bold{g}(\bm{a}) =\left(\frac{\partial ll}{\partial a_1}, \ldots, \frac{\partial ll}{\partial a_n}\right)^T,\]
\[\bm{H}(\bm{a}) = \left(\left( \frac{\partial^2 ll}{\partial a_j a_r} \right)\right)_{j=1:n, r=1:n}.\]
The Newton-Raphson iterations are carried out as follows:
\begin{itemize}
\item[] (1) Begin with starting values for $\bm{a}^{(0)}$;
\item[] (2) At any stage $k$, compute $\bm{a}^{(k+1)}= \bm{a}^{(k)}- H^{-1}\cdot \bold{g}(\bm{a}^{(k)})$;
\item[] (3) Repeat step (2) until convergence, i.e $\|\bm{a}^{(k+1)}-\bm{a}^{(k)} \|<\epsilon$.
\end{itemize}
\newpage
\subsection*{A.4 Bias Correction in Bootstrap Sampling}
\begin{figure}[ht]
    \centering
\caption{\label{fig:biascorr} Bias correction for bootstrap sampling of parameters. }
    \subfigure[Bootstrap Sampled Values for $(a_1,\ldots, a_{10})$]{\label{fig:a}\includegraphics[width=70mm]{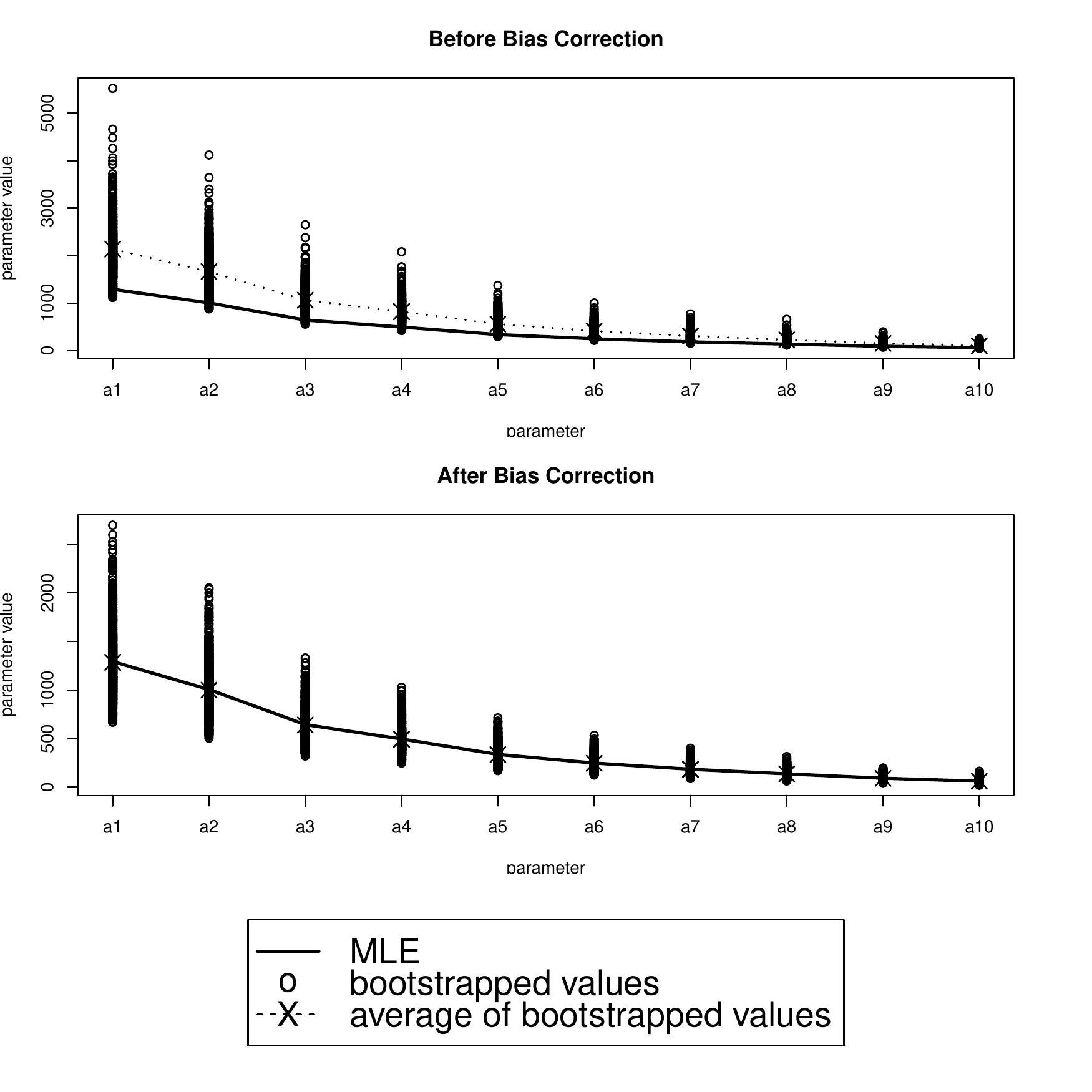}}
\subfigure[Bootstrap Sampled Values  for $(\phi_9,\ldots, \phi_{18})$]{\label{fig:b}\includegraphics[width=70mm]{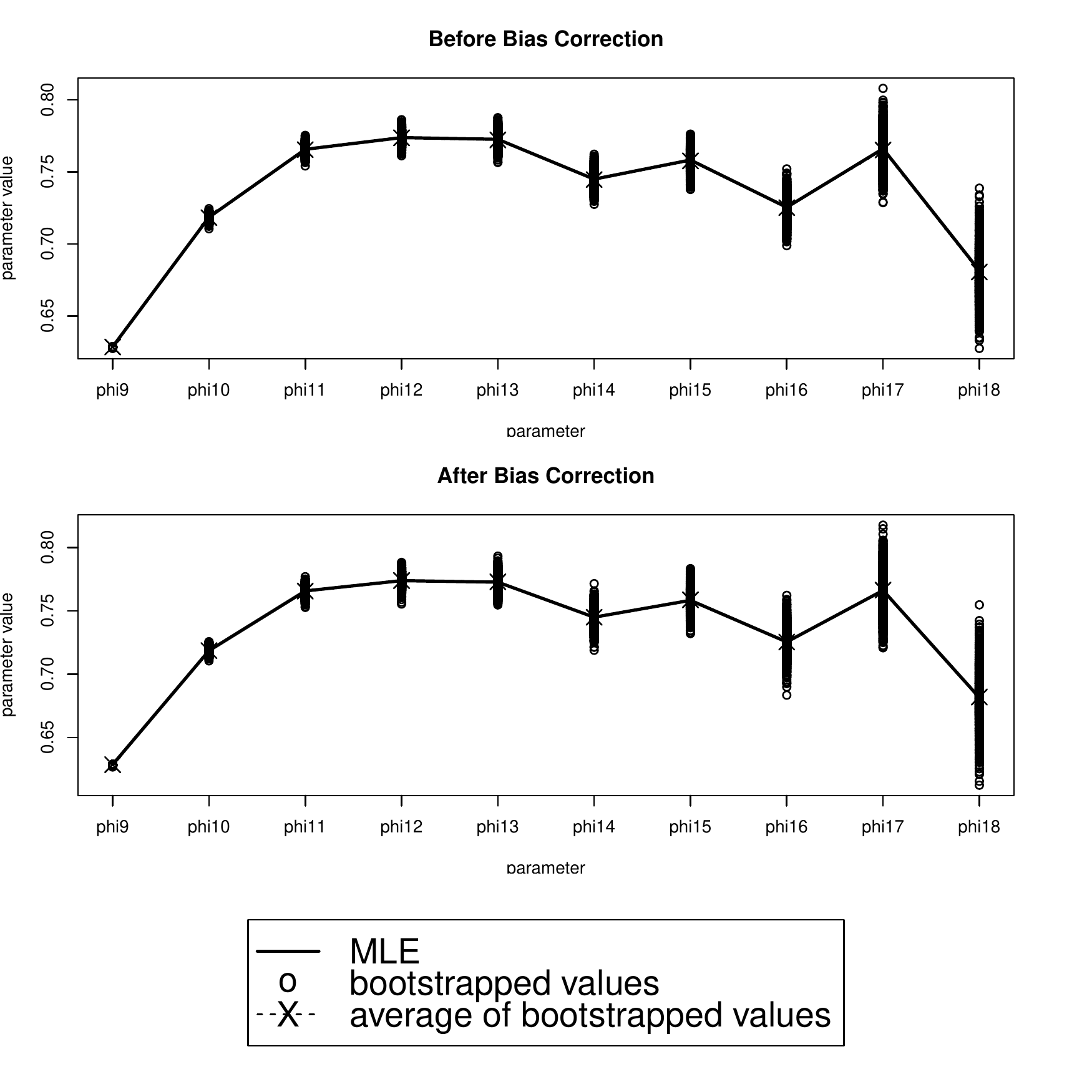}}
\end{figure}

In Section \ref{subsec:mle}, we described the steps for bootstrap for $\widehat{\bm{\theta}}^{MLE}$ and note the need for correcting the bias in the bootstrap samples. Here, provide a computational approach, based on a two-stage bootstrap procedure to correct for the bias in the bootstrap samples.
The first stage implements the same steps (1)- (4) of bootstrap as described in Section \ref{subsec:mle}. Then we determine a scaling factor for the MLE so that a repeated bootstrap with the scaled MLE, will result in parameter values with reduced bias. So,the second stage repeats the bootstrap procedure (1) - (4), but with the scaled MLE used in place of the original MLE. Specifically, we use the following steps:
\begin{itemize}
\item[I.] Using $\hat{\bm{\theta}}^{MLE}$, we obtain bootstrap samples for the vector ${\bm{\theta}}$ following steps (1)-(4) of Section \ref{subsec:mle}.
\item[II.] Calculate the average of the sampled vector, denoted by $\widehat{\bm{\theta}}^{avg}$. Then compute the modified MLE by scaling the original one:
\begin{align*}
& \widehat{\bm{\theta}}^{mod}=  \widehat{\bm{\theta}}^{MLE} \times \frac{\widehat{\bm{\theta}}^{MLE}}{\widehat{\bm{\theta}}^{avg}}.
\end{align*}
\item[III.] Repeat the bootstrap sampling steps (1)-(4) with $\widehat{\bm{\theta}}^{mod}$ instead of $\widehat{\bm{\theta}}^{MLE}$.
\end{itemize}
In our case, we observe that the bootstrap samples of $a_i$ show an upward bias in relation to the MLE, and the bootstrap samples of $\phi_i$ show a downward bias in relation to the MLE, although the bias in $\phi_i$ is less prominent.  Figure \ref{fig:biascorr} exhibits the bootstrap samples of parameters $a_j$ ($j=1,\ldots,10$) and $\phi_i$ ($i=1,\ldots,10$), before and after the bias correction. The results suggest that after the bias correction, the bootstrap sample mean matches closely with the MLE.

\end{document}